\documentclass[journal]{IEEEtran}
\usepackage{cite}
\usepackage{amsmath}

\usepackage{array}
\usepackage{url}
\usepackage{amssymb}
\usepackage{bm}
\usepackage{makecell}
\usepackage{overpic}    
\usepackage{booktabs}
\usepackage{multirow}
\usepackage{threeparttable}
\usepackage{color}
\allowdisplaybreaks
\usepackage{subfigure}
\usepackage[ruled,linesnumbered]{algorithm2e}

\begin{document}
\title{Emergency-Aware and Frequency-Constrained HVDC Planning for A Multi-Area Asynchronously Interconnected Grid}
\author{Yiliu~He,~\IEEEmembership{Student~Member,~IEEE,}
        Haiwang~Zhong,~\IEEEmembership{Senior~Member,~IEEE,}
        Grant~Ruan,~\IEEEmembership{Member,~IEEE,}
        Yan~Xu,~\IEEEmembership{Senior~Member,~IEEE,}
        and~Chongqing~Kang,~\IEEEmembership{Fellow,~IEEE}
\thanks{This work was supported by the National Natural Science Foundation of China under Grant U22B6007 and No. 72242105. \textit{(Corresponding author: Haiwang Zhong)}}
\thanks{Yiliu He, Haiwang Zhong, and Chongqing Kang are with the State Key Laboratory of Power System Operation and Control, Dept. of Electrical Engineering, Tsinghua University, Beijing, China.}
\thanks{Grant Ruan is with the Lab for Information \& Decision Systems, Massachusetts Institute of Technology, Massachusetts, U.S.}
\thanks{Yan Xu is with the Center for Power Engineering (CPE), School of Electrical and Electronic Engineering, Nanyang Technological University, Singapore.}
}

\maketitle
\begin{abstract}
High-voltage direct current (HVDC) technology has played a crucial role for long-distance transmission of renewable power generation.
However, the integration of large-capacity HVDC lines introduces significant frequency security challenges during HVDC fault emergencies.
This paper proposes an emergency-aware and frequency-constrained HVDC planning method to optimize the capacity of inter-area HVDC tie-lines in a multi-area asynchronously interconnected grid.
Firstly, a coordinated emergency frequency control scheme is proposed to allocate the emergency control resources during HVDC faults.
Then, an enhanced system frequency response model integrating event-driven emergency frequency control is developed and a weighted oblique decision tree approach is employed to extract frequency nadir security constraints.
The proposed planning model considers all potential HVDC fault emergencies while treating candidate HVDC capacities as decision variables.
Simulation results demonstrate superior performance in balancing economic efficiency with frequency security requirements, providing a practical solution for inter-area HVDC planning.

\end{abstract}


\section{Introduction}
\IEEEPARstart{H}{igh}-voltage direct current (HVDC) transmission technology is predominantly employed in long-distance bulk power transmission due to its economic efficiency and high controllability.
Many countries worldwide have initiated large-scale renewable energy projects.
Currently, most concentrated renewable energy resources, such as offshore wind farms and desert-based solar power plants, are located far from major load centers.
In such cases, HVDC has emerged as a preferred solution for inter-area renewable energy transmission and grid interconnection.
In China, more than 20 HVDC and Ultra-HVDC lines are currently in operation to transmit electricity from remote renewable power plants to load centers \cite{sun2017renewable}.
According to \cite{entso2020completing}, nearly 30\% of the European Union's projected renewable capacity installation will rely on inter-area HVDC links.

Despite its technical and economic benefits, the large scale integration of HVDC poses significant challenges to power system operation and stability control.
An HVDC line fault typically induces a more severe power imbalance in regional grids than any conventional generator tripping events.
As the capacity of individual HVDC lines continues to grow with technological advancements, their impact on the frequency security of the connected grids becomes increasingly critical.
During an HVDC fault, the receiving-end system experiences a serious frequency drop while the sending-end system faces over-frequency issues.
In extreme cases, under-frequency load shedding or over-frequency generator tripping may be triggered, resulting in substantial economic losses.
Moreover, the high penetration of converter-interfaced renewable generation further reduces the system inertia, which exacerbates the risk of large frequency deviations \cite{du2020frequency}.
Consequently, it is necessary to evaluate the impact of HVDC lines on system security for HVDC capacity planning and the allocation of frequency support resources.

Ensuring power system frequency security has been addressed through various optimization frameworks. In conventional power systems dominated by synchronous generators, research has focused on integrating frequency constraints into unit commitment problems. Reference \cite{ahmadi2014security} first proposed the embedding of piece-wise linear frequency nadir constraints in the unit commitment model, which has been further extended to frequency-constrained stochastic unit commitment by \cite{badesa2019simultaneous} and \cite{paturet2020stochastic}. Recent research has extended these concepts to incorporate converter-interfaced resources \cite{zhang2020modeling}. References \cite{chu2020towards} and \cite{zhang2022frequency} specifically addressed wind power integration, with \cite{chu2020towards} optimizing wind turbine synthetic inertia provision, and \cite{zhang2022frequency} developing a unit commitment model considering wind frequency support capabilities and wake effect. For HVDC interconnected systems, \cite{wen2017enhancing} proposed control strategies for enhancing frequency stability of asynchronous grids, while \cite{tosatto2021towards} and \cite{he2024multi} developed unit commitment models considering inter-area frequency support by HVDC. In planning contexts, researchers have embedded frequency security into generation and storage expansion problems. Reference \cite{zhang2021frequency} proposed a co-planning framework for generation and energy storage with frequency constraints. Reference \cite{li2021frequency} incorporated frequency response support from wind power in generation planning, while reference \cite{cao2022chance} focused on the configuration of battery energy storage systems. References \cite{wogrin2020assessing} further considered inertia and reactive power constraints in generation expansion planning.
However, the impact of large-capacity HVDC on system frequency security is seldom considered in the existing literature, and optimal HVDC capacity planning remains to be further studied.

To accurately model system frequency dynamics under large power imbalances, event-driven emergency frequency control (EFC) should be incorporated in the analysis, especially for inter-area HVDC faults.
Event-driven EFC serves as a critical corrective measure for large disturbances, where conventional frequency regulation (e.g., generator primary frequency control) may be insufficient in both response speed and magnitude to maintain frequency within secure limits.
Unlike conventional measures, event-driven EFC can be instantaneously activated by fault detection signals, delivering predetermined power responses within a short time. 
The primary resources for event-driven EFC include \cite{wang2019real, shi2022coordinating}:
\begin{itemize}
    \item Direct load control (DLC): Enables immediate reduction of pre-contracted, non-essential loads during emergencies, providing faster and more targeted response than under-frequency load shedding (UFLS).
    \item HVDC emergency power control (EPC): Facilitates inter-area frequency support among asynchronously interconnected areas by re-modulating HVDC power flow to mitigate power imbalances.
\end{itemize}
The specific control schemes for event-driven EFC, including resource selection and power set-points, can be pre-optimized and automatically deployed during real-time operation.
With rising renewable penetration and decreasing system inertia, the event-driven EFC has become indispensable for frequency security during major disturbances. 
Consequently, it is necessary to explicitly consider event-driven EFC measures for the accurate assessment of system frequency response during HVDC fault emergencies.
Although previous works have studied the EFC of individual synchronous system, the coordination strategies of EFC resources among asynchronously interconnected areas has received little attention and requires further investigation.

To address the existing research gap, this paper proposes an emergency-aware and frequency-constrained HVDC planning method that optimizes inter-area HVDC capacities while ensuring system frequency security. 
The main contributions of this work are summarized as follows:
\begin{enumerate}
    \item A coordinated emergency frequency control scheme for a multi-area asynchronously interconnected grid is proposed to mitigate large power imbalances caused by HVDC faults.
    The scheme optimally allocates the EPC capabilities of remaining HVDC links and the DLC resources to enable frequency support among asynchronously interconnected areas.
    \item An enhanced SFR model is developed to consider event-driven EFC and characterize the frequency dynamics of a multi-area asynchronously interconnected grid during HVDC faults. Frequency security constraints are extracted using a weighted oblique decision tree method, enabling efficient evaluation of frequency nadir limits. 
    \item The emergency-aware and frequency-constrained HVDC planning is formulated as a stochastic optimization model with a tri-layer structure: the planning layer for capacity decisions, the operation layer for scenario-based scheduling, and the emergency control layer for EFC resource allocation.
    This structure ensures systematic evaluation of all potential HVDC faults across diverse operation scenarios.    
\end{enumerate}

The remainder of this paper is organized as follows: Section \ref{control_scheme} presents the coordinated EFC scheme. Section \ref{sfr} develops the enhanced SFR model and describes the frequency nadir constraint extraction methodology. Section \ref{optim_model} formulates the planning model. Section \ref{case_study} analyzes case results and Section \ref{conclusion} provides conclusions.

\section{Coordinated Emergency Frequency Control}
\label{control_scheme}

\subsection{Event-Driven Emergency Frequency Control}
Event-driven EFC is specifically designed to address large power disturbances through rapid response measures.
Unlike response-based frequency control methods (under-frequency load shedding and over-frequency generator tripping) that are activated only when the frequency reaches critical thresholds \cite{xie2021distributional}, event-driven EFC initiates corrective actions immediately upon fault detection \cite{wang2011event}.
This proactive approach yields significantly lower system impact and implementation costs compared to conventional response-based measures.

In this study, we utilize HVDC EPC to enable mutual frequency support across interconnected areas. When an emergency is detected, a control signal is transmitted to the HVDC modulation controller, which adjusts the active power setpoint to a predetermined value according to the specific emergency. 
The HVDC EPC response can be modeled as a delayed step function \cite{wang2020coordinated}:
\begin{equation}
    \Delta P_{EPC, l}(t)= u\left(t-\tau_{EPC}\right)\Delta P_{EPC, l}
    \label{2a11}
\end{equation}
where $u(\cdot)$ denotes the unit step function, $\tau_{EPC}$ is the control time delay, and $\Delta P_{EPC, l}$ is the EPC response magnitude of HVDC line $l$.
Because HVDC systems respond rapidly to power setpoint adjustments and they are under direct control of the system operator, the value of $\tau_{EPC}$ is very small (typically around 100 ms\cite{wang2020coordinated,wang2019real}).

Similarly, the response power of DLC can be modeled as a delayed step function:
\begin{equation}
    \Delta P_{DLC}(t)= u\left(t-\tau_{DLC}\right)\Delta P_{DLC}  
    \label{2a12}
\end{equation}
where $\tau_{DLC}$ is the time delay and $\Delta P_{DLC}$ is the response magnitude of DLC.
Due to communication delay and load characteristics, the time delay of direct load control $\tau_{DLC}$ is generally longer than that of HVDC EPC.

\subsection{Coordination Strategy for Asynchronously Interconnected Areas}
\label{section_coordinated_control}
Adjustment of HVDC power setpoint can affect both sides of the HVDC line, so, HVDC EPC is essentially the mutual frequency support among the areas interconnected by HVDC.
This bidirectional control capability allows areas experiencing power imbalances to receive support from the other interconnected area during emergencies.
By coordinating the control of the response magnitude of HVDC EPC and allocating frequency regulation reserves among asynchronously interconnected areas, the system can maintain frequency security across all interconnected areas during emergencies while minimizing the operational costs.

\begin{figure}[htbp]
    \centering
    \includegraphics[width=0.48\textwidth]{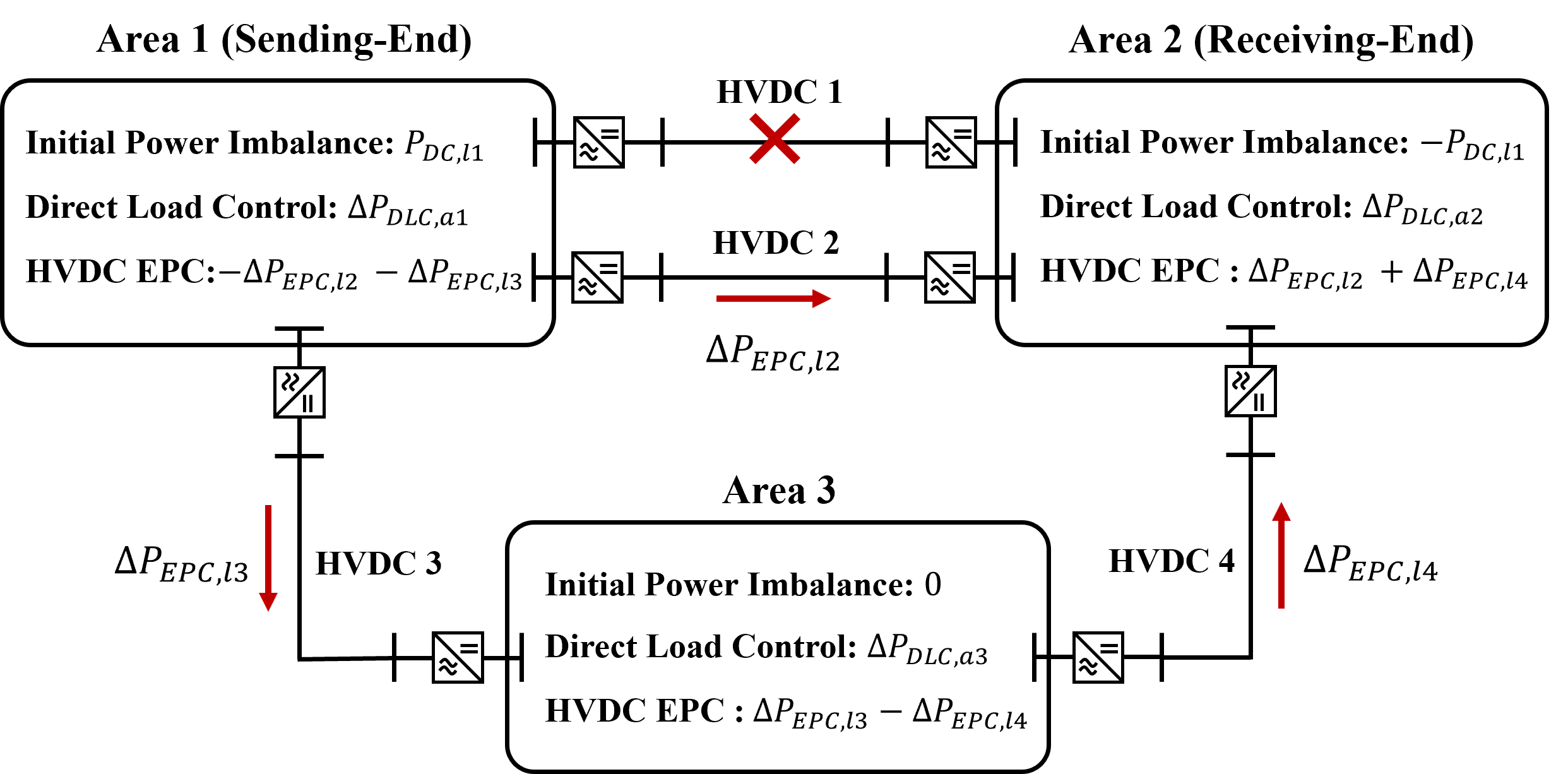}
    \caption{Coordinated EFC scheme for a multi-area asynchronously interconnected grid.}
    \label{coordinated_control}
\end{figure}

The coordinated EFC scheme aims to allocate the response power of remaining HVDC lines and DLC resources across all areas while ensuring frequency security.
For each potential emergency, the coordinated EFC scheme is pre-determined and activated once the emergency occurs.
Fig. \ref{coordinated_control} qualitatively illustrates the mechanism of the coordinated EFC scheme for a multi-area asynchronously interconnected grid.
Consider a fault in HVDC 1, which imposes a power surplus of $P_{DC,l1}$ on Area 1 as well as a power shortage of $-P_{DC,l1}$ on Area 2. 
Here $P_{DC,l1}$ denotes the pre-fault active power flow on HVDC 1.
After detecting the emergency, the EPC of the remaining HVDC lines is triggered to provide mutual frequency support among these areas, mitigating the power imbalances in Area 1 and Area 2.
HVDC 2 (parallel to the faulted line) increases the power flow from Area 1 to Area 2 by $\Delta P_{EPC,l2}$, directly compensating the imbalances in both areas.
HVDC 3 increases the power flow from Area 1 to Area 3 by $\Delta P_{EPC,l3}$, alleviating the surplus in Area 1.
Similarly, HVDC 4 increases the power flow from Area 3 to Area 2 by $\Delta P_{EPC,l4}$ to mitigate the shortage in Area 2.
Concurrently, pre-optimized DLC is triggered in each area to further suppress the frequency deviations.

The mathematical formulation of the optimal coordinated EFC scheme is given by \eqref{co_ctrl}, with the objective function \eqref{2b1} minimizing the total control costs of DLC and HVDC EPC.
Constraints \eqref{2b2} and \eqref{2b3} define the maximum available resources for DLC and HVDC EPC respectively, where $\mathcal{A}$ denotes the set of all asynchronous areas and $\mathcal{L}_{DC}$ denotes the set of all inter-area HVDC lines.
The net response power of HVDC EPC of the area $\Delta P_{EPC, a}$ is calculated as the sum of response powers from all connected HVDC lines, as expressed in \eqref{2b4}.
Here, $I_{DC,l}(a)$ is the connection indicator between HVDC line $l$ and Area $a$, with $I_{DC,l}(a)=-1$ if HVDC line $l$ originates from Area $a$, $I_{DC,l}(a)=1$ if HVDC line $l$ terminates in Area $a$, and $I_{DC,l}(a)=0$ if HVDC line $l$ is not connected to Area $a$.

Constraint \eqref{2b5} ensures frequency nadir security for each area, where $\Delta f_a(\Delta P_{D,a},\Delta P_{DLC, a},\Delta P_{EPC, a})$ denotes the expression of the frequency nadir given initial power imbalance $\Delta P_{D,a}$, DLC response power $\Delta P_{DLC, a}$, and the net response power of HVDC EPC $\Delta P_{EPC, a}$.
\begin{subequations}
\begin{align}
\min  \quad& \sum_{a \in \mathcal{A}} C_{DLC} \Delta P_{DLC, a}+\sum_{l \in \mathcal{L}_{DC}} C_{EPC} \Delta P_{EPC, l}\label{2b1} \\
s.t.\quad & \Delta P_{DLC, a}\leq\Delta  P_{DLC, a}^{\rm max}, \forall a \in \mathcal{A} \label{2b2}\\
& \Delta P_{EPC, l}\leq \Delta P_{EPC, l}^{\rm max}, \forall l \in \mathcal{L}_{DC}\label{2b3}\\
& \Delta P_{EPC, a} = \sum_{l\in \mathcal{L}_{DC}} I_{DC,l}(a) \Delta P_{EPC, l}, \forall a \in \mathcal{A}\label{2b4}\\
& \Delta f_a(\Delta P_{D,a},\Delta P_{DLC, a},\Delta P_{EPC, a})\leq \Delta f^{\rm max}_a, \forall a \in \mathcal{A}\label{2b5}
\end{align}
\label{co_ctrl}
\end{subequations}

\section{Frequency Nadir Constraint Extraction}
\label{sfr}
Determining the optimal coordinated EFC scheme for a given emergency requires accurate evaluation of the system frequency nadir specified in \eqref{2b5}.
This section develops an enhanced SFR model incorporating event-driven EFC to characterize the frequency dynamics.
To embed the non-linear frequency nadir constraint into the optimization model, we employ a weighted oblique decision tree approach to extract linear approximations from the dataset generated by the enhanced SFR model.

\subsection{Enhanced SFR Model with Event-Driven EFC}

This study focuses on the frequency support among asynchronously interconnected areas while ignoring nodal frequency variations within individual synchronous areas.
For each synchronous area, we adopt a multi-machine SFR model that assumes a uniform frequency across the entire area. This approach fundamentally accounts for the exclusive inertia and independent frequency evolution of each asynchronously interconnected area. The model does not aggregate the dynamics of different areas into a single system but rather treats each area as a separate entity with its own dynamic response, interconnected only through the power setpoints of the HVDC links.
Consequently, the frequency dynamics of each area are governed exclusively by its own internal power imbalance, system inertia, and governor responses, as defined by its unique swing equation. This allows for the accurate simulation of the distinct frequency evolution processes across the multi-area grid following a disturbance.

The linear and time-invariant nature of the SFR model allows direct superposition of event-driven EFC with the initial power imbalance.
Consequently, the total power imbalance $\Delta P_a(s)$ can be expressed as \eqref{3a_16}, where $\Delta P_{D,a}$ is the initial power imbalance, $\Delta P_{DLC,a}$ is the DLC power, $\tau_{DLC}$ is the time delay of DLC, $\Delta P_{EPC,a}$ is the net response power of HVDC EPC, and $\tau_{EPC}$ is the time delay of HVDC EPC.
\begin{equation}
     \Delta P_a(s) = \frac{\Delta P_{D,a}}{s} + \frac{\Delta P_{DLC,a} e^{-s\tau_{DLC}}}{s} + \frac{\Delta P_{EPC,a} e^{-s\tau_{EPC}}}{s}
    \label{3a_16}
\end{equation}

The system frequency response can be characterized  by the swing equation \eqref{3a_11}, where $\Delta P_a(s)$ is the total power imbalance, subscript $a$ denotes Area $a$, $H_a$ is the system inertia, $D_a^{\rm load}$ is the load damping ratio, and $\Delta f_a(s)$ is the system frequency deviation.
$G_{T,k}(s)$, $G_{H,k}(s)$, and $G_{E,k}(s)$ represent the response characteristics of thermal units, hydro units, and energy storage systems, respectively.
The subscript $k$ denotes the index of the unit.
The load damping ratio $D_a^{\rm load}$ is typically proportional to the total system load.
We assume that the system inertia is provided by thermal and hydro units, as shown in \eqref{3a_12}, where $u_k$ is the commitment status of thermal unit $k$.
\begin{align}
    \nonumber
    \Delta P_a(s)=&\left(2H_a s+D^{\rm load}_a+
    \sum_{k\in \mathcal{T}_a} G_{T,k}(s)\right. \\
   & \left.
    + \sum_{k\in \mathcal{H}_a} G_{H,k}(s)
    + \sum_{k\in \mathcal{E}_a} G_{E,k}(s)
    \right)\Delta f_a(s)
    \label{3a_11}
\end{align}
\begin{equation}
    H_a=\sum_{k\in \mathcal{T}_a}{H_k P_k^{\rm max}u_k}
    + \sum_{k\in \mathcal{H}_a}{H_k P_k^{\rm max}}
    \label{3a_12}
\end{equation}

The response characteristics of thermal units are given by \eqref{3a_13}, where $R_k$ is the governor regulation coefficient, $T_{G,k}$ is the governor time constant, $T_{C,k}$ is the steam chest time constant, $T_{R,k}$ is the reheat time constant, and $F_{H,k}$ is the high-pressure turbine fraction.
When the thermal generator is offline, $G_{T,k}(s)=0$, meaning it contributes no frequency regulation capability.
\begin{equation}
     G_{T,k}(s) = \frac{1+F_{H,k}T_{R,k}s}{R_k\left(1+T_{G,k}s\right) \left(1+T_{C,k}s\right) \left(1+T_{R,k}s\right)}u_k
     \label{3a_13}
\end{equation}

The response characteristics of hydro units, accounting for water hammer effects, are given by \eqref{3a_14}, where $R_{P,k}$ is the permanent droop, $R_{T,k}$ is the temporary droop, $T_{G,k}$ is the governor time constant, $T_{R,k}$ is the rest time, and $T_{W,k}$ is the water starting time constant that represents the penstock dynamics \cite{fernandez2022frequency}.
\begin{equation}
     G_{H,k}(s) = \frac{T_{R,k}s+1}{R_{P,k}\left(1+T_{G,k}s\right)\left(\frac{R_{T,k}}{R_{P,k}}T_{R,k}s+1\right)}
 \frac{1-T_{W,k}s}{1+\frac{T_{W,k}}{2}s}
      \label{3a_14}
\end{equation}

For the energy storage systems, we assume that droop frequency control is applied, as shown in \eqref{3a_15}, where $R_{E,k}$ is the droop regulation coefficient and $T_{E,k}$ is the control delay time constant.
\begin{equation}
     G_{E,k}(s) = \frac{1}{R_{E,k}\left(1+T_{E,k}s\right)}
           \label{3a_15}
\end{equation}

\subsection{Dimensionality Reduction by Equivalent Parameter Aggregation}

To account for the variation of system operational state, the frequency nadir constraint \eqref{2b5} is reformulated as \eqref{3b1}, where $\bm{S}_a = \left[u_1,\dots, u_{K_a},Load_a\right]$ is the system operational state variables of Area $a$, including the commitment status of thermal generators and the system load.
\begin{equation}
    \Delta f_a\left(\Delta P_{D,a},\Delta P_{DLC, a},\Delta P_{EPC, a},\bm{S}_a\right)\leq \Delta f^{\rm max}_a
    \label{3b1}
\end{equation}

Direct utilization of $\bm{S}_a$ as the input variable for frequency nadir constraint expression would lead to dimensionality challenges in dataset generation and fitting, due to the large quantity of generators.
To address this, we employ equivalent aggregated parameters to represent the system operational state.
Proposed by \cite{shi2018analytical}, the equivalent parameter aggregation method is effective to simplify the multi-machine SFR model into a single-machine SFR model while preserving accuracy.
We modify the original formulations from \cite{shi2018analytical} to better accommodate variable load damping conditions.
The three equivalent aggregated parameters in this work are the equivalent system inertia $H_a$, the fast-acting equivalent droop $D^{\rm fast}_a$, and the slow-acting equivalent droop $D^{\rm slow}_a$.
The expression of the equivalent system inertia is given by \eqref{3a_12}.
The fast-acting equivalent droop $D^{\rm fast}_a$ represents the system equivalent droop with very small time constant, as given by \eqref{3b_12}, consisting of load damping characteristics, high-pressure turbines of thermal units, hydro units, and energy storage systems.
The slow-acting equivalent droop $D^{\rm slow}_a$ represents the system equivalent droop with relatively large time constant, as given by \eqref{3b_13}, mainly from low-pressure turbines of thermal units.
By introducing these equivalent aggregated parameters, the frequency nadir constraint \eqref{3b1} can be simplified into \eqref{3b14}.
Let $\bm{X_a}=[H_a, D^{\rm fast}_a, D^{\rm slow}_a, \Delta P_{EPC, a}, \Delta P_{DLC, a},\Delta P_{D,a}]$ denote the input variable vector.
The goal of linear frequency nadir constraint extraction is to find a set of linear hyperplanes in the form of \eqref{3b15} that represent the secure region defined by \eqref{3b14}.
\begin{equation}
    D^{\rm fast}_{a}=D^{\rm load}_a + \sum_{k\in \mathcal{T}_a}\frac{F_{H,k}P_k^{\rm max}u_k}{R_k}
    + \sum_{k\in \mathcal{H}_a}\frac{ P_k^{\rm max}}{R_k}  
    + \sum_{k\in \mathcal{E}_a}\frac{ P_k^{\rm max}}{R_k} 
    \label{3b_12}
\end{equation}
\begin{equation}
    D^{\rm slow}_a=\sum_{k\in \mathcal{T}_a}\frac{\left(1-F_{H,k}\right)P_k^{\rm max}u_k}{R_k}
    \label{3b_13}
\end{equation}
\begin{equation}
    \Delta f_a\left(\bm{X_a}\right)\leq \Delta f^{\rm max}_a
    \label{3b14}
\end{equation}
\begin{equation}
    \bm{A}\bm{X_a}+\bm{b}\le \bm{0}
    \label{3b15}
\end{equation}

It is noteworthy that the equivalent parameters in this study are formulated in absolute physical values (e.g., MW·s for inertia) rather than being normalized by the system load. This formulation is selected for its distinct advantages in a linear constraint extraction context. First, it provides a direct physical representation of the total frequency support resources, ensuring dimensional consistency with the absolute power imbalances and emergency control actions (MW) modeled in the optimization. Second, and critically, it ensures robustness across different load conditions. Normalizing equivalent parameters by the time-varying system load would result in significant fluctuations and a widely dispersed feature space. Such a large parameter range tends to increase the nonlinearity of the security boundary, making it difficult to approximate with linear constraints. By using absolute values, the feature space remains more compact and determined primarily by physical unit commitment; this stability facilitates the WODT in extracting accurate linear frequency constraints that are valid across diverse operational scenarios.

The adoption of equivalent aggregated parameters significantly enhances computational tractability for the constraint extraction process. It is crucial to clarify that this aggregation serves solely as a dimensionality reduction technique to define the input feature space for the machine learning model. The dataset generation process relies on the high-fidelity, multi-machine SFR model, ensuring that the simulated frequency dynamics accurately reflect detailed generator responses and EFC actions. Consequently, the approximation lies only in the mapping from the complex system states to the simplified feature vector used for constraint learning.

While this mapping introduces a degree of simplification, its impact on the accuracy of the final constraints is minimal for two reasons. First, as validated in \cite{shi2018analytical}, these aggregated parameters are derived to preserve the dominant dynamic characteristics of the system, with reported errors in frequency nadir prediction typically remaining below 2\%. Second, and more importantly, the subsequent WODT training process inherently compensates for residual approximation errors. By learning optimal linear security boundaries directly from the high-fidelity simulation results, the WODT effectively calibrates the constraints within the aggregated feature space, ensuring a reliable security assessment.

\subsection{Simulation-Based SFR Dataset Generation}
\label{section:dataset_generation}
\begin{figure}[htbp]
    \centering
    \includegraphics[width=0.46\textwidth]{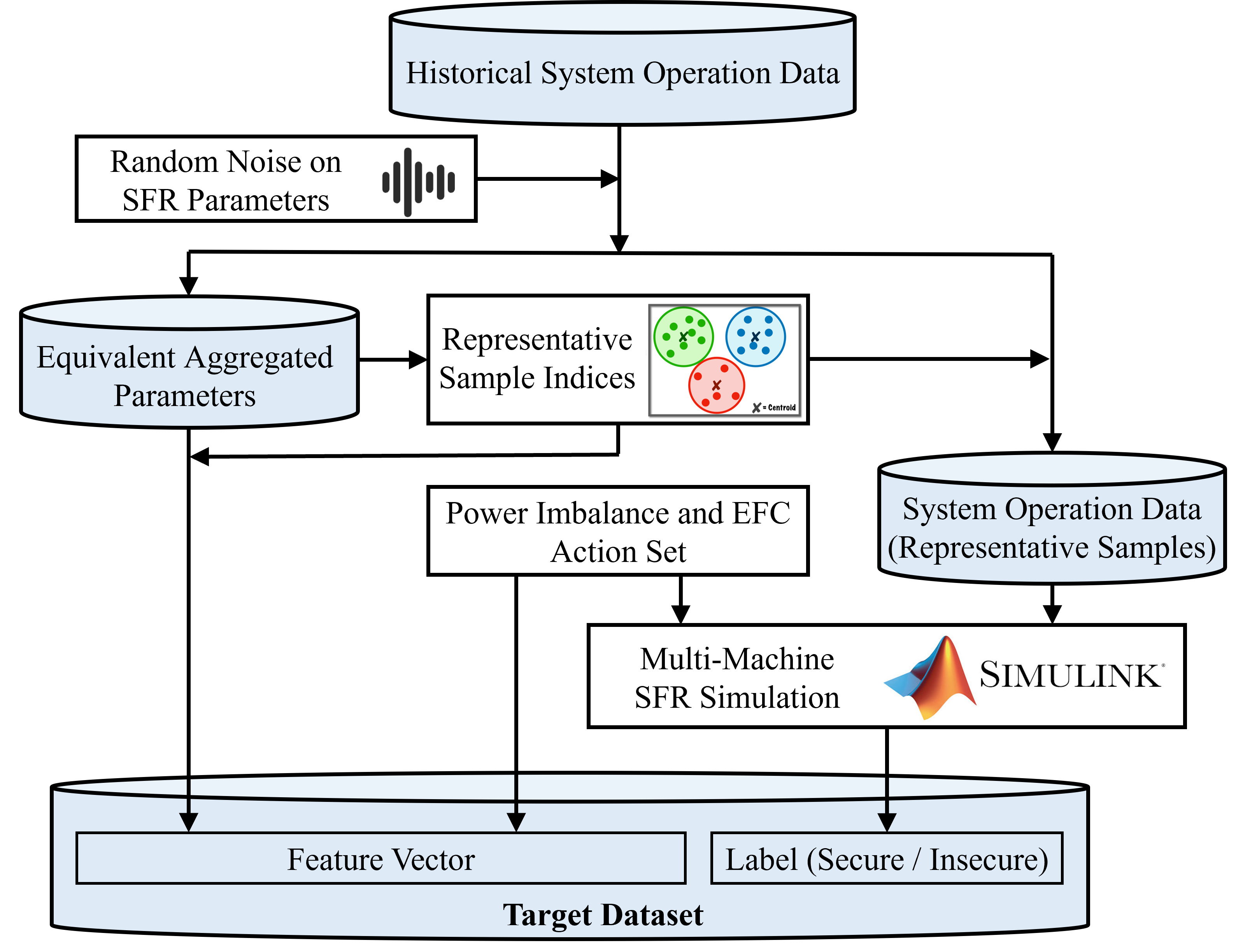}
    \caption{Process for simulation-based SFR dataset generation.}
    \label{fig-dataset}
\end{figure}

A simulation-based approach is adopted for SFR dataset generation for two main reasons. First, unlike simplified SFR models that use analytical expressions with idealized assumptions, our simulation approach explicitly incorporates practical governor-turbine dynamics of different generation types, including thermal reheat cycles, hydro water hammer effects, and converter-based responses. This ensures accurate capture of frequency evolution characteristics that are crucial for reliable security assessment but cannot be represented in simplified analytical models. Second, simulation enables the creation of a comprehensive dataset covering diverse operational scenarios, including extreme contingencies and high renewable penetration levels. These scenarios may be unavailable in historical records but are essential for robust long-term planning.

Dataset plays an essential role in training machine learning models.
To generate high-quality dataset and extract linear frequency nadir constraints with high accuracy and generalization ability, we proposed a simulation-based SFR dataset generation process, which is illustrated in Fig. \ref{fig-dataset}.
We assume that the historical system operation data are available, including the commitment status of generators and the total load.
In this work, the historical system operation data are obtained by implementing conventional unit commitment model with year-round data of hourly loads and renewable energy outputs.

To enhance dataset diversity and simulate parameter uncertainties, we introduce random perturbations to the principal SFR parameters. Specifically, the inertia constant $H_k$, the governor regulation coefficient $R_k$ for all types of units, and the high-pressure turbine fraction $F_{H,k}$ for thermal units are perturbed by $\pm 50\%$ relative to their nominal values using a uniform distribution. This process significantly increases the diversity of the training data, thereby improving the generalization capability of the trained WODT model. Consequently, the extracted security constraints exhibit greater robustness against the parameter variations and measurement uncertainties encountered in real-world systems. The equivalent aggregated parameters are subsequently calculated based on these perturbed values according to \eqref{3a_12}, \eqref{3b_12}, and \eqref{3b_13}.

To address potential redundancy in historical system operation data, we perform clustering on the equivalent aggregated parameters to identify representative samples. 
The indices of these representative samples are recorded, based on which a subset of the most distinctive historical system operation states is filtered.
Moreover, a comprehensive set of possible power imbalances and various EFC actions are incorporated as the simulation input.
Detailed multi-machine SFR simulations are implemented for each combination of the system operation status, the power imbalance, and the EFC actions.
The maximum frequency deviation can be obtained from the simulation results, and a label is attached to the data point based on whether the maximum frequency deviation exceeds the security threshold.
The resulting feature vector of the data point comprises the equivalent aggregated parameters, the power imbalance, and the EFC actions.

\subsection{Rule Extraction by Weighted Oblique Decision Tree}

The weighted oblique decision tree (WODT) \cite{yang2019weighted} is selected to linearize the frequency nadir constraints due to its distinct advantages in the context of large-scale planning optimization. First, it demonstrates superior scalability in handling high-dimensional datasets compared to methods like adaptive piece-wise linear fitting, which often struggle to efficiently partition multi-dimensional feature spaces without incurring excessive computational complexity. Second, the output of the WODT is inherently compatible with Mixed-Integer Linear Programming (MILP) frameworks. The learned model consists of polyhedral regions defined by linear hyperplanes, which can be directly embedded into the optimization problem using binary variables. Finally, WODT offers enhanced interpretability, and the learned hyperplanes define physically meaningful security boundaries within the feature space. This approach maintains high accuracy in capturing nonlinear security boundaries while ensuring the mathematical tractability required for the planning model.

The scalability of the WODT method to large datasets and varying data ranges is a key advantage. Computationally, the algorithm efficiently processes large-scale training data by employing gradient-based optimization on a differentiable objective function, allowing it to distill complex patterns from millions of samples into a compact tree structure without the combinatorial explosion often seen in other methods. Furthermore, the WODT is robust to varying data ranges and feature scales. Its core mechanism of oblique splitting enables the construction of optimal hyperplanes that separate data based on linear combinations of all features, effectively capturing coupled relationships regardless of absolute magnitudes. To further stabilize the training process and enhance generalization across diverse operating ranges, a feature standardization step is applied to normalize all input variables prior to training.

Let $\boldsymbol{a}_t$ denote the split parameter of node $t$.
The sigmoid-based weights of a data point belonging to the right child node and left child node are given by \eqref{3b21} and \eqref{3b22} respectively, where $\boldsymbol{X}_{(n)}$ is the feature vector of the $n^{\rm th}$ sample.
The total weights in left and right child nodes are calculated using \eqref{3b24}.
For data points with label $k$, we further denote the label-specific weight sums in left and right child nodes as $W_L^k$ and $W_R^k$, which can be calculated via \eqref{3b25}.
\begin{equation}
w_{(n)}^R=\sigma(\boldsymbol{a}_t^T\boldsymbol{X}_{(n)})=\frac{1}{1+e^{-\boldsymbol{a}_t^T\boldsymbol{X}_{(n)}}}
    \label{3b21}
\end{equation}
\begin{equation}
   w_{(n)}^L = 1- w_{(n)}^R=\frac{e^{-\boldsymbol{a}_t^T\boldsymbol{X}_{(n)}}}{1+e^{-\boldsymbol{a}_t^T\boldsymbol{X}_{(n)}}}
    \label{3b22}
\end{equation}
\begin{equation}
   W_L=\sum w_{(n)}^L,\quad W_R=\sum w_{(n)}^R
    \label{3b24}
\end{equation}
\begin{equation}
   W_L^k=\sum_{y_{(n)}=k}w_{(n)}^L,\quad W_R^k=\sum_{y_{(n)}=k}w_{(n)}^R
    \label{3b25}
\end{equation}

The weighted information entropies of the left and right child nodes can be calculated by \eqref{3b26} and \eqref{3b27} respectively.
After splitting the node, the total weighted information entropy is given by \eqref{3b28}, which serves as the objective function for the splitting process.
\begin{equation}
   E_L = -\sum_k \frac{W_L^k}{W_L}\log_2\frac{W_L^k}{W_L}
    \label{3b26}
\end{equation}
\begin{equation}
   E_R = -\sum_k \frac{W_R^k}{W_R}\log_2\frac{W_R^k}{W_R}
    \label{3b27}
\end{equation}
\begin{equation}
   E=W_L E_L + W_R E_R
    \label{3b28}
\end{equation}

The objective function $E$ is continuous and differentiable with respect to the split parameter $\boldsymbol{a}_t$. 
Therefore, the problem of minimizing the objective function $E$ can be solved by unconstrained non-linear optimization methods, such as the quasi-Newton method.
The node splitting process repeats iteratively until either the maximum tree depth is reached or the node purity satisfies the stopping criterion.
Following the training of the weighted oblique decision tree, each leaf node is assigned a label based on the data points it contains.

To explicitly formulate the frequency nadir constraints, we define the feature vector $X_a$ for Area $a$ as $\bm{X_a}=[H_a, D^{\rm fast}_a, D^{\rm slow}_a, \Delta P_{EPC, a}, \Delta P_{DLC, a},\Delta P_{D,a}]$, which includes the initial power imbalance, the response power of emergency controls, and the equivalent aggregated SFR parameters. The WODT partitions the high-dimensional feature space into multiple polyhedral regions, each corresponding to a leaf node. 

Let $\mathcal{T}_{secure}$ denote the set of leaf nodes classified as secure. For a specific secure leaf node $t \in \mathcal{T}_{secure}$, the feasible region is defined by the intersection of the splitting hyperplanes along the path from the root node to leaf $t$. This can be expressed as a set of linear inequalities in the form of \eqref{3b31}, where $\bm{A}_t$ and $\bm{b}_t$ are coefficient matrices determined by the training process. 
Specifically, for a leaf with depth$=D$, the corresponding coefficient matrices are $\bm{A}_t=\left[\bm{A}_t^1;\dots; \bm{A}_t^D \right]$ and $\bm{b}_t=\left[b_t^1;\dots; b_t^D \right]$, where $\bm{A}_t^d$ and $b_t^d$ are separating coefficients at depth$=d$.
Since the system is secure if the operating point falls into any of the secure leaf regions, the frequency security constraint is the union of these polytopes. 
To incorporate the rules of multiple leaf nodes into the mixed-integer linear programming (MILP) formulation, we employ the Big-M relaxation method as shown in \eqref{3b33}. Binary variables $v_t$ are introduced to indicate which specific secure leaf node contains the current operating point, where $v_t=1$ indicates the rule is active and $v_t=0$ indicates the rule is inactive.
The constraint \eqref{3b32} ensures that exactly one rule remains active at any time by requiring that only one binary variable $v_t$ equals 1 while all others equal 0.

\begin{equation}
   \bm{A}_t\bm{X}_a+\bm{b}_t\geq \bm{0}
    \label{3b31}
\end{equation}
\begin{equation}
   \bm{A}_t\bm{X}_a+\bm{b}_t\geq-M(1-v_t)*\bm{1}
    \label{3b33}
\end{equation}
\begin{equation}
   \sum_t v_t=1,v_t\in\{0,1\}
    \label{3b32}
\end{equation}

The WODT-extracted constraints demonstrate strong generalization capabilities suitable for evolving power systems. This adaptability is achieved by the use of equivalent aggregated parameters as the feature space. Since the constraints are learned based on total system physical properties (e.g., total inertia and damping) rather than specific unit commitments, the addition or retirement of generators can be directly accommodated by recalculating these aggregated values, without invalidating the learned decision boundaries. Furthermore, the training process, which covers a wide spectrum of parameter variations and emergency scenarios, ensures the model remains valid across diverse operating conditions. For long-term application, the entire workflow supports periodic updates: as the power system evolves significantly, the dataset can be regenerated and the constraints can be retrained to ensure the planning model continuously reflects the latest system characteristics.

\section{Emergency-Aware HVDC Planning Method}
\label{optim_model}
\begin{figure}[htbp]
    \centering
    \includegraphics[width=0.48\textwidth]{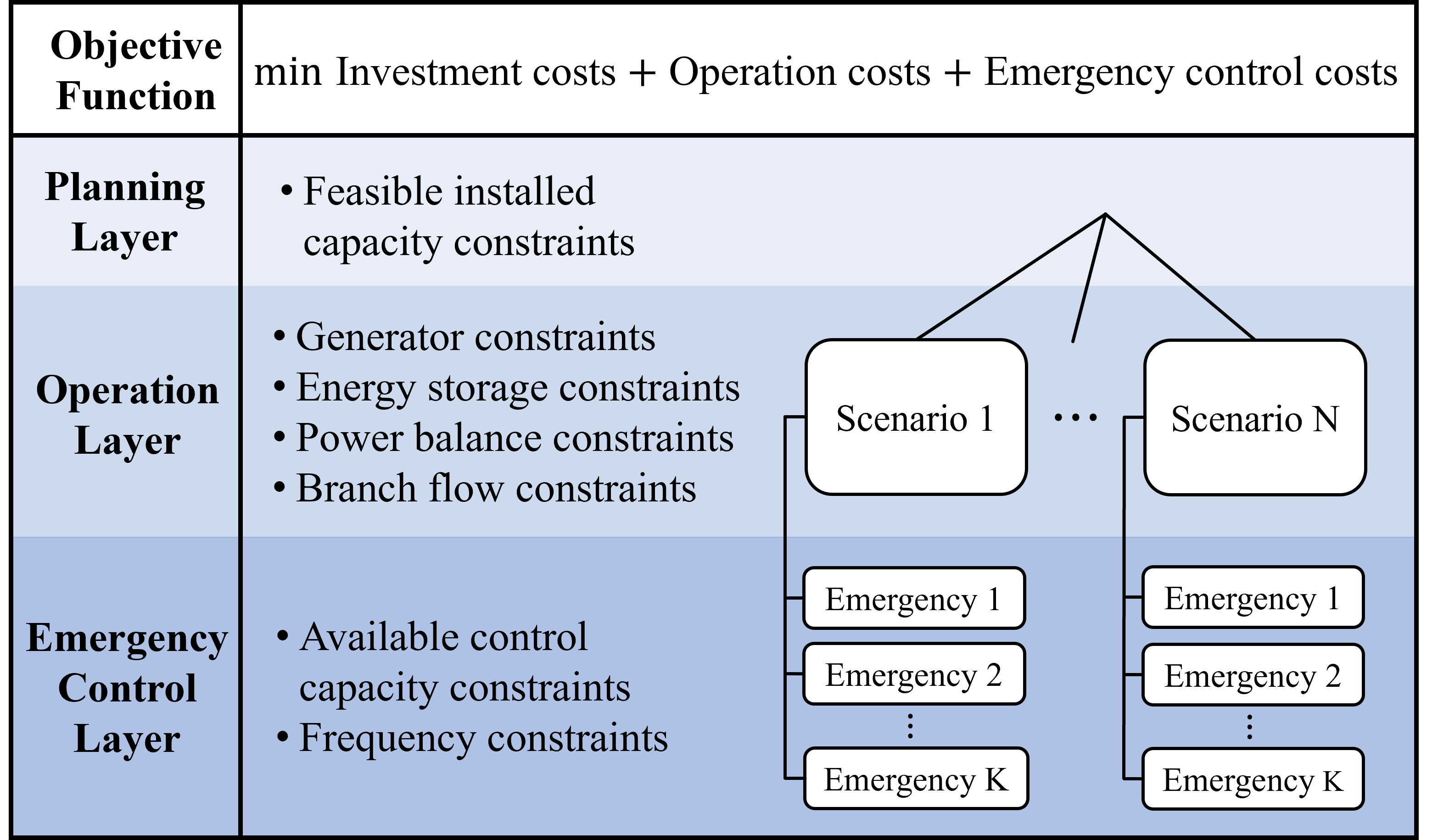}
    \caption{The framework of the emergency-aware HVDC planning model.}
    \label{fig:framework}
\end{figure}

This section formulates the emergency-aware HVDC planning as a stochastic programming problem. The overall model framework is illustrated in Fig. \ref{fig:framework}.
The objective function minimizes three cost components: the annualized investment costs $C_{ INV}^{ ANN}$, the annualized operation costs $C_{ OP}^{ ANN}$, and the annualized emergency control costs $C_{ EC}^{ ANN}$, as shown in \eqref{1a1}.
\begin{equation}
\min C_{ INV}^{ ANN} + C_{ OP}^{ ANN} + C_{ EC}^{ ANN}
\label{1a1}
\end{equation}
In the planning layer, the feasible installed capacities constraints for candidate HVDC links are considered. 
Candidate energy storage systems are also incorporated for potential deployment to address operational needs.
In the operation layer, the unit commitment model under representative scenarios is implemented, enforcing generator constraints, energy storage system constraints, power balance constraints, and branch flow constraints.
In the emergency control layer, the emergency control resources for all potential emergencies are allocated, incorporating the control capacity limits and frequency security constraints.

\label{optim_model_structure}
\subsection{Planning Layer}
The annualized investment cost of the planning model is given by \eqref{1a11}, where $r$ is the discount factor, $N$ is the device lifespan, $z_k^{ESS}$ is the binary variable indicating the installation of energy storage system $k$, $C_{INV,k}^{ESS}$ is the investment cost of energy storage system $k$, $\mathcal{E}^c$ is the set of candidate energy storage systems, $C_{{ INV},l}^{ DC}$ is the investment cost of HVDC line $l$, and $\mathcal{L}^c_{DC}$ is the set of candidate HVDC lines.
\begin{equation}
C_{ INV}^{ ANN}=\frac{r}{1-(1+r)^{-N}}\left(\sum_{k\in \mathcal{E}^c}{z_k^{ESS}C_{INV,k}^{ESS}}+\sum_{l\in \mathcal{L}^{c}_{DC}}C_{{ INV},l}^{ DC}\right)
\label{1a11}
\end{equation}

As established in \cite{hartel2017review}, the HVDC investment cost consists of the transmission line investment $C_{{ INV},l}^{ L}\left(L_l,P_{l}^{ CAP}\right)$ and the converter station investment $C_{{ INV},l}^{ C}\left(P_{l}^{ CAP}\right)$, where $L_l$ is the length of HVDC line $l$, and $P_{l}^{ CAP}$ is the installed capacity of HVDC line $l$.
These cost components are specified in \eqref{1a3} and \eqref{1a4}, where $B_{Lp}$, $B_L$, $B_0$, $N_p$, $N_0$ are the cost parameters and $z_l^{ DC}$ is the binary variable indicating the installation of HVDC line $l$.

\begin{align}
&
C_{{ INV},l}^{ L}\left(L_l,P_{l}^{ CAP}\right)=B_{Lp}L_lP_{l}^{ CAP}+(B_LL_l+B_0)z_l^{ DC}
\label{1a3}
\\
&
C_{{ INV},l}^{ C}\left(P_{l}^{ CAP}\right)=N_pP_{l}^{ CAP}+N_0z_l^{ DC}
\label{1a4}
\end{align}

We assume that the length of each candidate HVDC line is fixed, then the total investment cost of the HVDC line can be given by the linear expression in \eqref{1a6}, where $C_l^{FIX}$ is the fixed cost component and $C_l^{CAP}$ is the capacity-dependent cost.
\begin{equation}
    C_{{ INV},l}^{ DC}=z_l^{DC}C_l^{FIX}+P_{l}^{ CAP}C_l^{CAP}
    \label{1a6}
\end{equation}

Due to limited equipment types, the capacities of HVDC lines are generally selected from several discrete values.
Therefore, discrete sizing constraints are enforced for candidate HVDC lines, as shown in \eqref{1b4} and \eqref{1b41}, where $P_{\rm min}^{ CAP}$ is the minimum installable capacity, $P_{\rm max}^{ CAP}$ is the maximum installable capacity, $\Delta P^{ CAP}$ is the capacity increment, and $n_l^{DC}$ is a non-negative integer variable to decide the installed capacity.
\begin{equation}
P_{l}^{ CAP} = z_l^{DC}P_{\rm min}^{ CAP}+n_l^{DC}\Delta P^{ CAP}
    \label{1b4}
\end{equation}
\begin{equation}
P_{l}^{ CAP} \leq z_l^{DC}P_{\rm max}^{ CAP}
    \label{1b41}
\end{equation}
\subsection{Operation Layer}
The annual operational cost is given by \eqref{1a7}, where $\Omega_{S}$ is the set of representative scenarios, $\pi_s$ is the weight of scenario $s$, $\Omega_{T}$ is the set of time periods within each scenario, $\mathcal{T}$ is the set of thermal generators, $C_k\left(P_k^{s,t}\right)$ is the piecewise-linear production cost of generator $k$, $v^{s,t}_k,w^{s,t}_k$ are binary variables indicating start-up and shutdown actions, and $C^{SU}_k,C^{SD}_k$ are the start-up and shutdown costs.
The superscript $s,t$ identifies scenario $s$ and time period $t$.
\begin{equation}
    C_{OP}^{ ANN} =\sum_{s\in\Omega_{S}}\pi_s\sum_{t\in\Omega_{T}}\sum_{k\in\mathcal{T}}\left(C_k\left(P_k^{s,t}\right)+C^{SU}_kv^{s,t}_k+C^{SD}_kw^{s,t}_k\right)
    \label{1a7}
\end{equation}

The active power output limits of thermal generators are defined by \eqref{4b_31}, where $P_k^{\rm min},P_k^{\rm max}$ are the minimum and maximum active power of generator $k$, and the 0-1 variable $u_k^{s,t}$ is the binary commitment status variable of generator $g$ in scenario $s$ at time $t$.
The upward and downward ramping constraints are defined by \eqref{4b_32} and \eqref{4b_33}, where $P_k^{\rm up}$ is the upward ramping limit and $P_k^{\rm down}$ is the downward ramping limit.
The minimum online/offline time constraints are defined by \eqref{4b_34} and \eqref{4b_36}, where $T_k^{\rm on}, T_k^{\rm off}$ are the minimum online and offline time requirements of generator $g$.
\begin{gather}
P_k^{\rm min}u_k^{s,t}\le P_k^{s,t}\le P_k^{\rm max}u_k^{s,t}
\label{4b_31}
\\
P_k^{s,t}-P_k^{s,t-1}\le P_k^{\rm up}v_k^{s,t}
\label{4b_32}
\\
P_k^{s,t-1}-P_k^{s,t}\le P_k^{\rm down}w_k^{s,t}
\label{4b_33}
\\
\sum_{\tau=t-T_k^{\rm on}{-1}}^{t-1}u_k^{s,\tau}\geq w_k^{s,t}T_k^{\rm on}
\label{4b_34}
\\
\sum_{\tau=t-T_k^{\rm off}{-1}}^{t-1}\left(1-u_k^{s,\tau}\right)\geq v_k^{s,t}T_k^{\rm off}
\label{4b_36}
\end{gather}

The energy storage systems follow the operational constraints given by \eqref{e1}-\eqref{e5}, where $e_k^{s,t}$ is the stored energy, $\eta_c,\eta_d$ are the charging and discharging efficiencies, $P_{c,k}^{s,t},P_{d,k}^{s,t}$ are the charging and discharging power, $\Delta t$ is the duration of each time period, $P_{c,k}^{\max},P_{d,k}^{\max}$ are the maximum charging and discharging power, $x_{c,k}^{s,t},x_{d,k}^{s,t}$ are the binary indicators for charging and discharging, $e_k^{\min},e_k^{\max}$ are the minimum and maximum storage capacities.
\begin{gather}
e_k^{s,t+1}=e_k^{s,t}+\eta_cP_{c,k}^{s,t}\Delta t-\frac{1}{\eta_d}P_{d,k}^{s,t}\Delta t
\label{e1}
\\
P_{c,k}^{s,t}\leq P_{c,k}^{\max}x_{c,k}^{s,t},\quad P_{d,k}^{s,t}\leq P_{d,k}^{\max}x_{d,k}^{s,t}
\label{e3}
\\
x_{c,k}^{s,t}+x_{d,k}^{s,t}\leq1
\label{e4}
\\
e_k^{\min}\leq e_k^{s,t}\leq e_k^{\max}
\label{e5}
\end{gather}
For the candidate energy storage systems, the operation is constrained by the installation state:
\begin{equation}
    x_{c,k}\leq z_k,\quad x_{d,k}\leq z_k
    \label{e6}
\end{equation}

The power balance constraints for AC network are shown in \eqref{4b_21}, where $P_{G,i}^{s,t}$ is the total power generation on bus $i$, $P_{D,i}^{s,t}$ is the total demand on bus $i$, $\theta_i^{s,t},\theta_j^{s,t}$ are the voltage phase angles on bus $i$ and $j$, $B_{i,j}$ is the equivalent susceptance between bus $i$ and $j$, $P_{DC,l}^{s,t}$ is the power injection of HVDC line $l$, and $I_{DC,l}(i)$ is the connection indicator between HVDC line $l$ and bus $i$. Here the losses of HVDC transmission are ignored.
\begin{equation}
P_{G,i}^{s,t}-P_{D,i}^{s,t}=\sum_{j}B_{i,j}\left(\theta_i^{s,t}-\theta_j^{s,t}\right)+\sum_{l\in \mathcal{L}_{DC}}P_{DC,l}^{s,t}I_{DC,l}(i)
\label{4b_21}
\end{equation}

The branch flow constraints for AC lines and DC lines are shown in \eqref{4b_41} and \eqref{4b_42} respectively, where $P_{AC,l}^{\rm max}$ is the capacity limit of AC transmission line $l$, $B_{l}$ is the AC line susceptance, $\mathcal{L}_{AC}\left(i,j\right)$ is the set of AC lines between bus $i$ and $j$.
$P_{DC,l}^{\rm max}$ is the capacity limit of DC transmission line $l$.
\begin{align}
&
-P_{AC,l}^{\rm max}\le B_{l}\left(\theta_i-\theta_j\right)\le P_{AC,l}^{\rm max},\forall l\in\mathcal{L}_{AC}\left(i,j\right)
\label{4b_41}
\\
&
-P_{DC,l}^{\rm max}\le P_{DC,l}^{s,t}\le P_{DC,l}^{\rm max},\forall l\in\mathcal{L}_{DC}\left(i,j\right)
\label{4b_42}
\end{align}
For the candidate HVDC line, the active power $P_{DC,l}^{s,t}$ is constrained by the installed capacity, as shown in \eqref{1b2}. 
\begin{equation}
-P_{l}^{ CAP}\le P_{DC,l}^{s,t}\le P_{l}^{ CAP}
    \label{1b2}
\end{equation}

\subsection{Emergency Control Layer}
For each potential HVDC fault emergency in set $\Omega_{E}$, the optimal coordinated EFC problem described in Section \ref{section_coordinated_control} is solved to determine the emergency response scheme.
The annual emergency control cost is given by \eqref{ec_cost}, where $\pi_e$ is the occurrence probability of emergency $e$, and $C_{E C}^{s, t, e}$ is the control cost under emergency $e$ at time $t$ in scenario $s$. The detailed expression of $C_{E C}^{s, t, e}$ has been defined in Section \ref{section_coordinated_control}.
\begin{equation}
C_{EC}^{ ANN}
=\sum_{s\in\Omega_{S}}\pi_s\sum_{t\in \Omega_{T} }\sum_{e\in\Omega_{E}}\pi_e C_{E C}^{s, t, e}
\label{ec_cost}
\end{equation}

The constraints of the available emergency control resources are given by \eqref{2b2}-\eqref{2b4}.
Linear frequency rules extracted by the weighted oblique decision tree in \eqref{3b31}-\eqref{3b32} are incorporated, replacing the original nonlinear frequency nadir constraints.

To ensure that the activation of emergency power control does not cause cascading overloads on the remaining healthy transmission lines, a post-disturbance branch flow constraint is enforced. This constraint ensures that the superposition of the pre-fault active power flow and the emergency power adjustment remains within the capacity limits of the HVDC lines, as shown in \eqref{dcpf}, where $P_{DC,l}^{s,t}$ represents the pre-fault power flow determined in the operation layer, $\Delta P_{EPC,l}^{s,t,e}$ is the emergency power adjustment of HVDC line $l$ under emergency $e$. This constraint guarantees that the mutual frequency support provided by the healthy HVDC lines does not violate their thermal or stability limits.

\begin{equation}
	-P_{DC,l}^{\rm max}\le P_{DC,l}^{s,t}+ \Delta P_{EPC, l}^{s,t,e}\le P_{DC,l}^{\rm max},\forall l\in\mathcal{L}_{DC}\left(i,j\right)
	\label{dcpf}
\end{equation}

\section{Case Study}
\label{case_study}
\subsection{Case Setup}
\begin{figure}[htbp]
\centering
\includegraphics[width=0.42\textwidth]{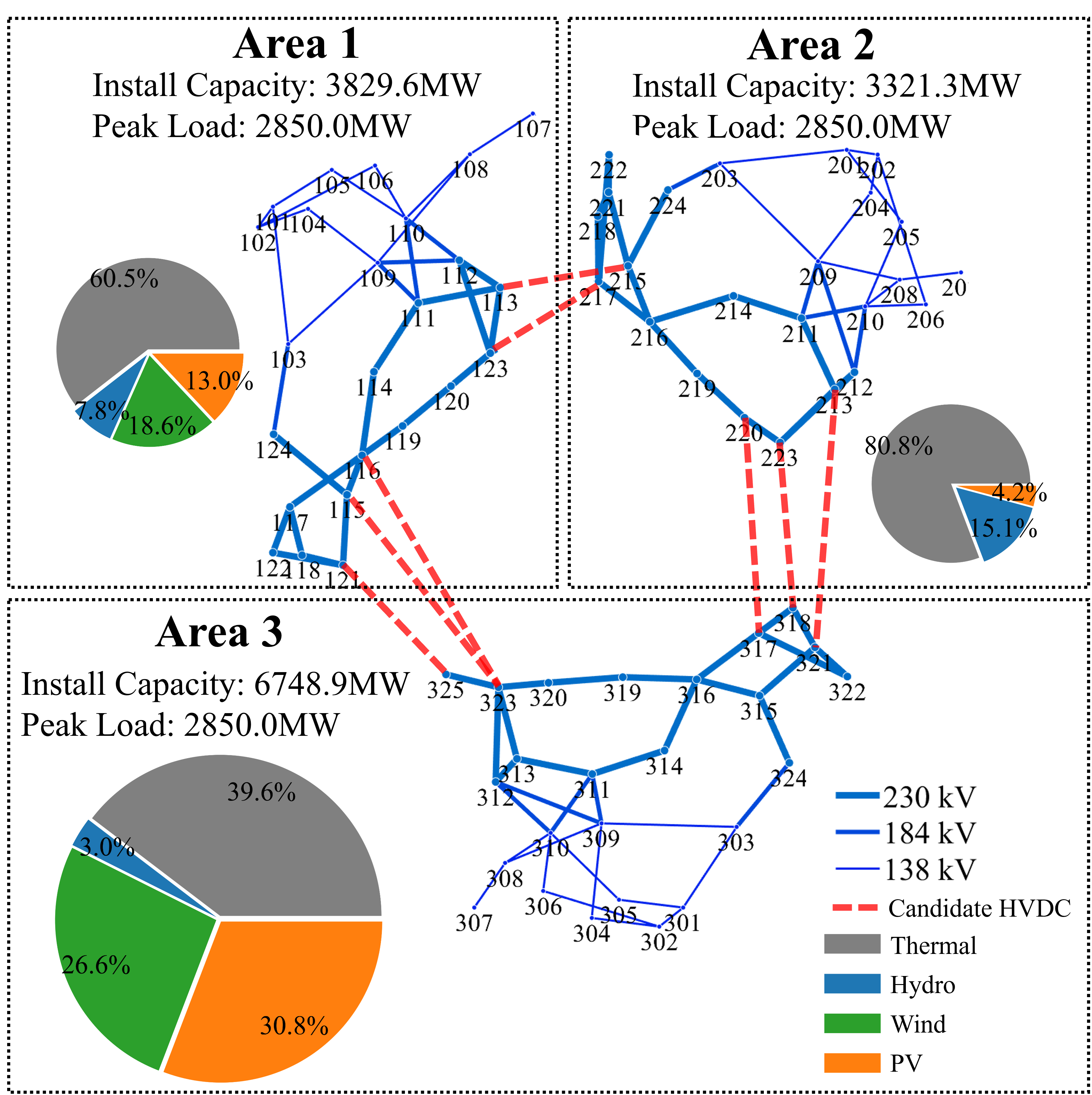}
\caption{The geographical layout and the generation mix of the modified RTS-GMLC test system.}
\label{testbed_fig}
\end{figure}
The three-area RTS-GMLC case \cite{barrows2019ieee} is adopted as the testbed system, which consists of 74 buses, 72 thermal units, 20 hydro units, 56 PV units, and 4 wind farms.
The original five inter-area tie lines are removed and replaced with candidate HVDC lines for inter-area power transmission.
Table \ref{candidate_hvdc} presents the candidate HVDC line specifications, while Fig. \ref{testbed_fig} shows the geographical layout and generation mix of the modified RTS-GMLC system.

Frequency regulation parameters are set as follows: 
\begin{itemize}
    \item For thermal generators, $R_k=0.06$, $F_{H,k}=0.3$, $T_{R,k}=12s$, $T_{G,k}=0.5s$, and $T_{C,k}=0.5s$.
    \item For hydro generators, $R_{P,k}=0.08$, $R_{T,k}=0.3$, $T_{G,k}=0.5s$, $T_{R,k}=12s$, $T_{W,k}=0.4s$.
    \item For energy storage systems, $R_{E,k}=0.05$, $T_{E,k}=0.5s$.
\end{itemize}
Emergency control delays are $\tau_{DLC}=0.6s$ for direct load control and $\tau_{EPC}=0.2s$ for HVDC EPC response. 
In each time period, the fault probability of each HVDC is 0.0001.
The control cost for HVDC EPC and DLC are \$100/MW and \$1000/MW respectively.
All AC areas have a load damping ratio $D^{load}_a=1$. Inertia constants come from the RTS-GMLC repository, and the frequency nadir security bound is 0.5Hz.

The planning model uses 8 representative scenarios selected by K-Medoids clustering, each with 24-hour time horizon at hourly resolution.
Nine candidate battery energy storage systems are located at buses 113, 115, 123, 216, 223, 211, 323, 316, and 311. Each storage system has 50MW power capacity, 100MWh energy capacity, 90\% charging/discharging efficiency, and \$1.96M annualized investment cost.
\begin{table}[htbp]
\centering
\caption{Information of the Candidate HVDC lines.}
\label{candidate_hvdc}
\begin{tabular}{cccccc}
\toprule
ID     & \begin{tabular}[c]{@{}c@{}}From\\ Bus\end{tabular} & \begin{tabular}[c]{@{}c@{}}To\\ Bus\end{tabular} & \begin{tabular}[c]{@{}c@{}}Length\\ (km)\end{tabular} & \begin{tabular}[c]{@{}c@{}}Annualized\\ Fixed\\ Cost (M\$)\end{tabular} & \begin{tabular}[c]{@{}c@{}}Annualized\\ Capacity\\ Cost (M\$/MW)\end{tabular} \\ \midrule
HVDC 1 & 113                                                & 215                                              & 105                                                   & 7.63                                                                    & 0.0156                                                                        \\
HVDC 2 & 123                                                & 217                                              & 100                                                   & 7.43                                                                    & 0.0153                                                                        \\
HVDC 3 & 318                                                & 223                                              & 120                                                   & 8.24                                                                    & 0.0165                                                                        \\
HVDC 4 & 317                                                & 220                                              & 125                                                   & 8.44                                                                    & 0.0168                                                                        \\
HVDC 5 & 321                                                & 213                                              & 130                                                   & 8.64                                                                    & 0.017                                                                         \\
HVDC 6 & 325                                                & 121                                              & 120                                                   & 8.24                                                                    & 0.0165                                                                        \\
HVDC 7 & 323                                                & 115                                              & 125                                                   & 8.44                                                                    & 0.0168                                                                        \\
HVDC 8 & 323                                                & 116                                              & 130                                                   & 8.64                                                                    & 0.017                                                                         \\ \bottomrule
\end{tabular}
\end{table}

\subsection{System Frequency Response Simulation with Event-Driven Emergency Control}
\begin{figure}[htbp]
    \centering
    \includegraphics[width=0.4\textwidth]{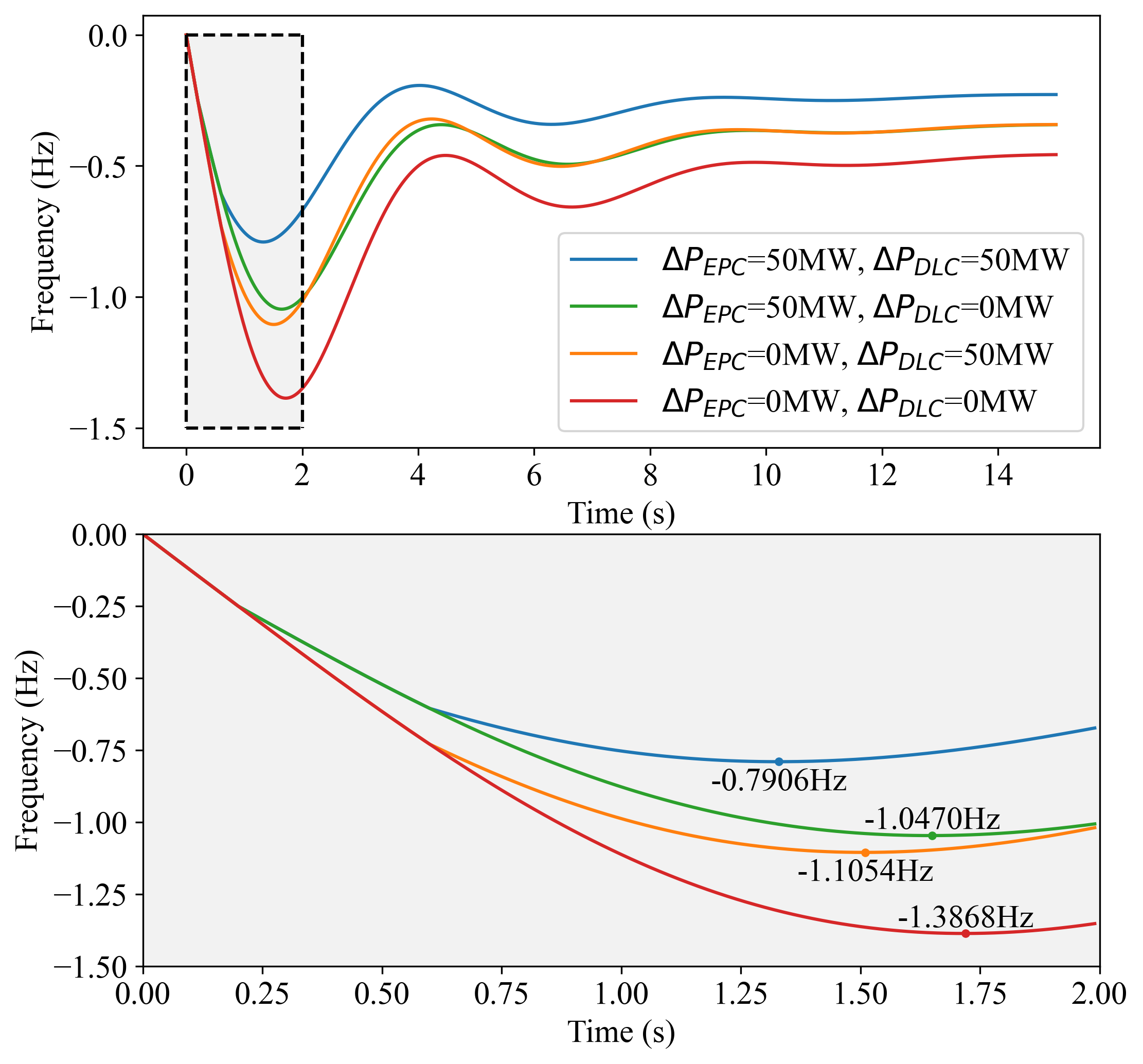}
    \caption{Simulation results of the system frequency response model with different emergency control schemes.}
    \label{fig:sfr_area1}
\end{figure}

To demonstrate the effectiveness of the emergency frequency control, we analyze the system response to a fixed power imbalance $\Delta P_{D,a}=200MW$ under different control configurations.
Fig. \ref{fig:sfr_area1} shows the simulated frequency response characteristics.
Without event-driven EFC activation, the system experiences a severe frequency deviation reaching 1.3868Hz. 
With $\Delta P_{DLC, a}=50MW$, the frequency decline is mitigated after the direct load control is activated ($t=0.6s$), reducing the maximum deviation to 1.1054Hz.
The HVDC emergency power control demonstrates superior performance due to its faster response time. With equivalent response power $\Delta P_{EPC, a}=50MW$, the maximum frequency deviation decreases further to 1.047Hz. 
This simulation validates the importance of response speed in emergency frequency control.
Notably, while the dynamic responses differ, both control methods achieve an identical quasi-steady-state frequency. 
The combined application of both control methods yields optimal results, limiting the maximum deviation to 0.7906 Hz.

To evaluate the robustness of the proposed method against communication and actuation uncertainties, a sensitivity analysis was performed on the time delays of the event-driven EFC. Table \ref{tab:sensitivity_delay} presents the maximum frequency deviation under the combined control scheme ($\Delta P_{EPC,a}=50$ MW, $\Delta P_{DLC,a}=50$ MW) with varying time delays for HVDC EPC ($\tau_{EPC}$) and DLC ($\tau_{DLC}$).

The results demonstrate a remarkable sensitivity to control latency. In the ideal scenario with zero delay, the maximum frequency deviation is limited to 0.6934Hz. However, as the delays increase to the base case ($\tau_{EPC}=0.2$s, $\tau_{DLC}=0.6$s), the deviation rises to 0.7906Hz. When the delays become excessive ($\tau_{EPC}=1.5$s, $\tau_{DLC}=2.0$s), the deviation reaches 1.3629Hz. This value approaches the uncontrolled scenario (1.3868Hz), indicating that if the control action occurs after the natural frequency nadir, the emergency response becomes ineffective. Therefore, ensuring low-latency communication is critical for the practical viability of the proposed scheme.

\begin{table}[htbp]
	\caption{Sensitivity Analysis of Frequency Nadir to EFC Time Delays}
	\label{tab:sensitivity_delay}
	\centering
	\begin{tabular}{ccc}
		\toprule
		HVDC EPC Delay & DLC Delay & Max. Frequency Deviation \\
		($\tau_{EPC}$) & ($\tau_{DLC}$) & (Hz) \\
		\midrule
			0 s & 0 s & 0.6934 \\
0.2 s & 0.6 s & 0.7906 \\
0.5 s & 1.0 s & 0.9735 \\
1.0 s & 1.5 s & 1.2111 \\
1.5 s & 2.0 s & 1.3629 \\
		\bottomrule
	\end{tabular}
\end{table}

\subsection{Dataset Generation and Frequency Constraint Extraction}

\begin{figure}[htbp]
\centering
\subfigure[Equivalent aggregated parameters of original dataset]{
        \includegraphics[width=0.21\textwidth]{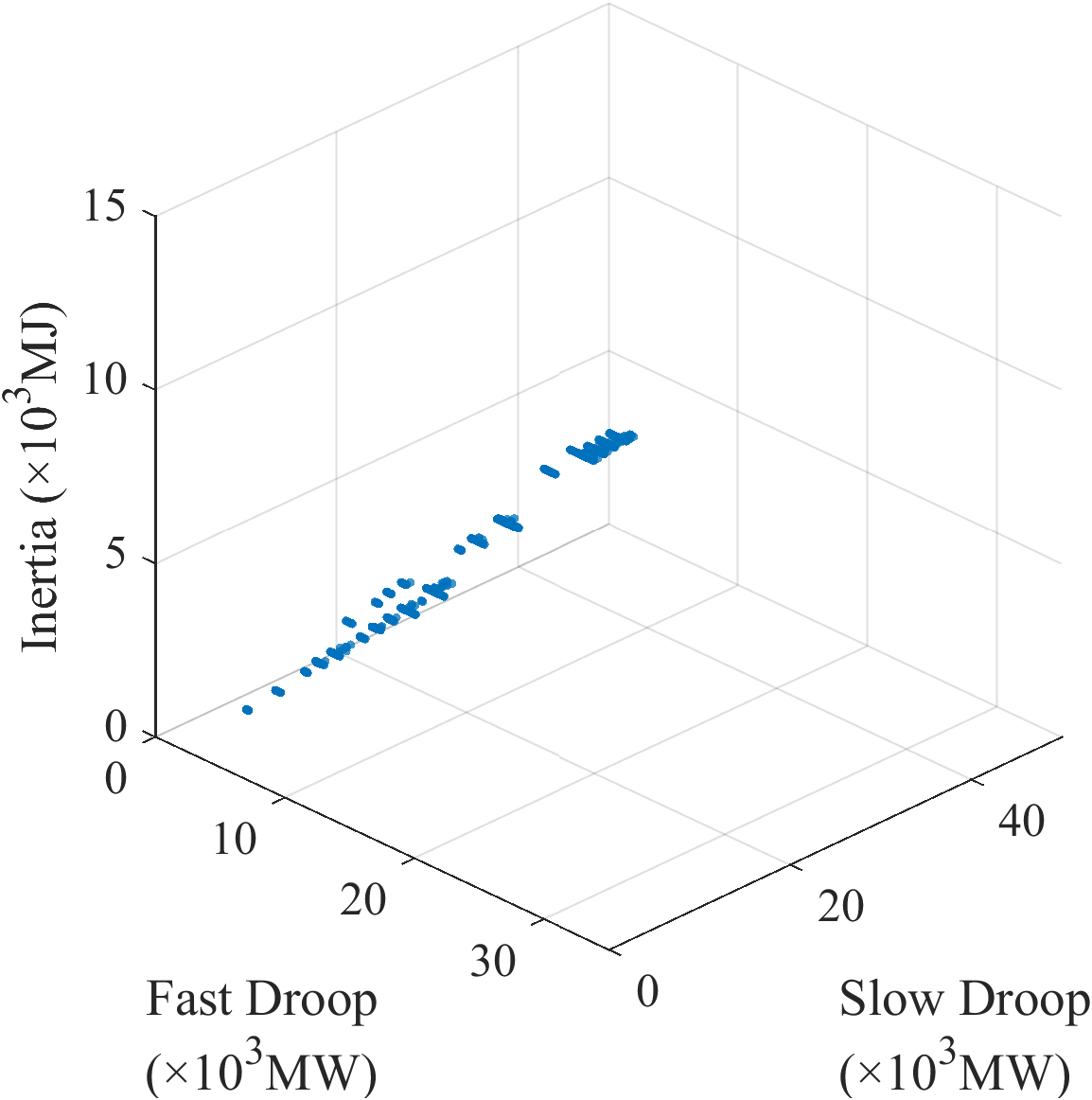}
        \label{fig:dataset0}}
\subfigure[Representative samples with random perturbations]{
        \includegraphics[width=0.21\textwidth]{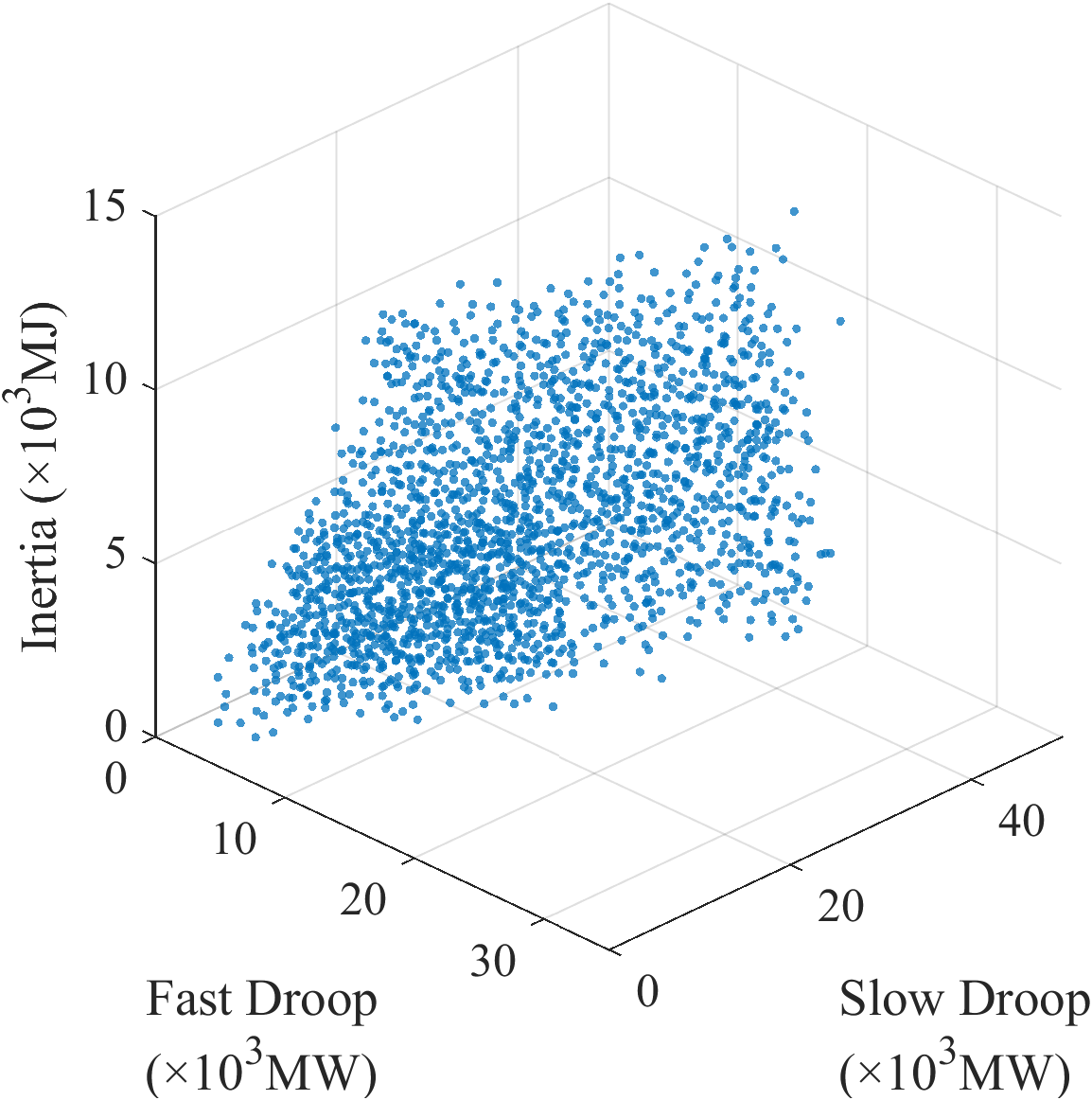}
        \label{fig:dataset1}}
\\
\subfigure[Three-dimensional view of the labeled dataset]{
        \includegraphics[width=0.21\textwidth]{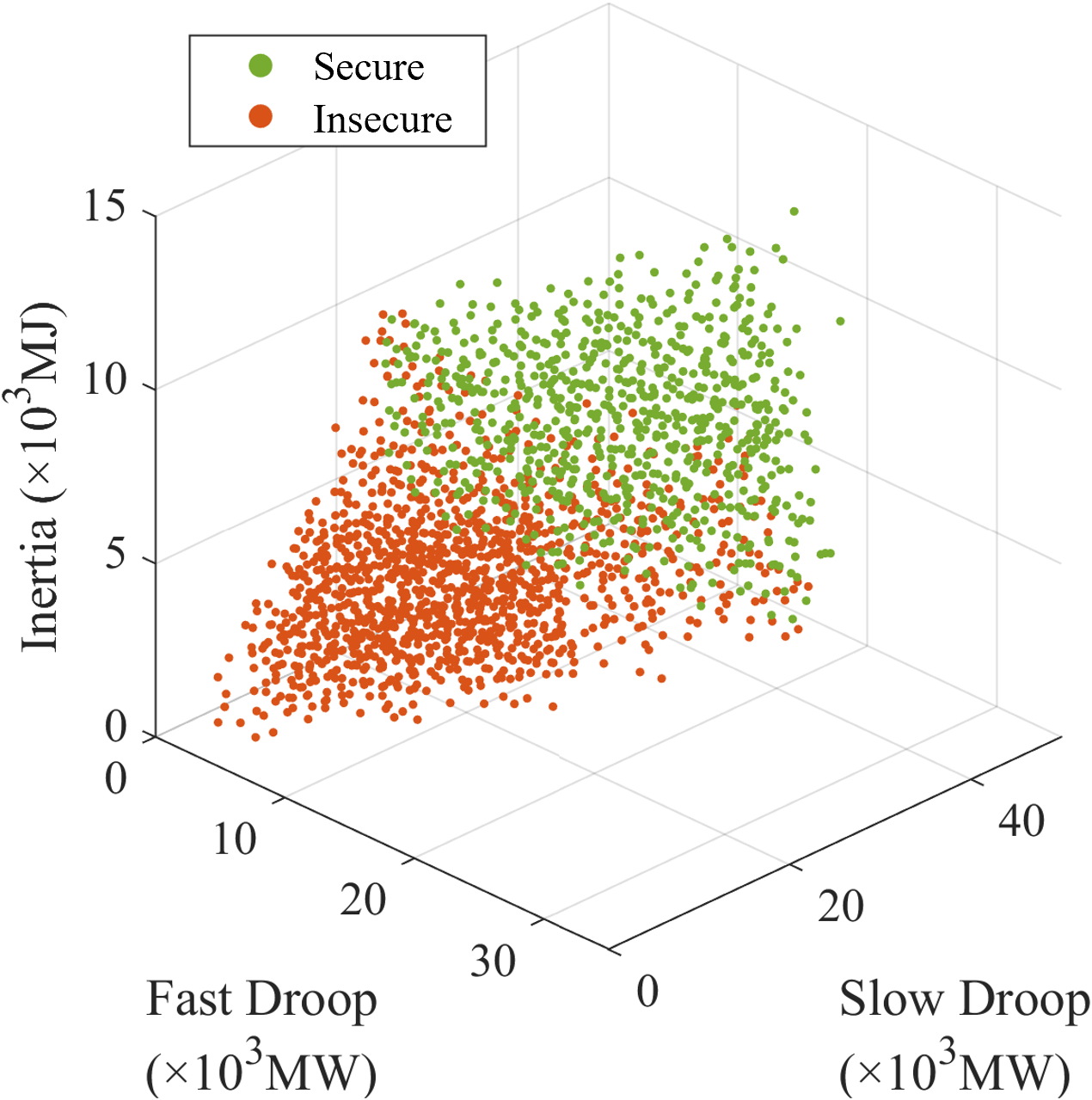}
        \label{fig:dataset2}}
\subfigure[Classification boundary of a secure leaf node]{
        \includegraphics[width=0.21\textwidth]{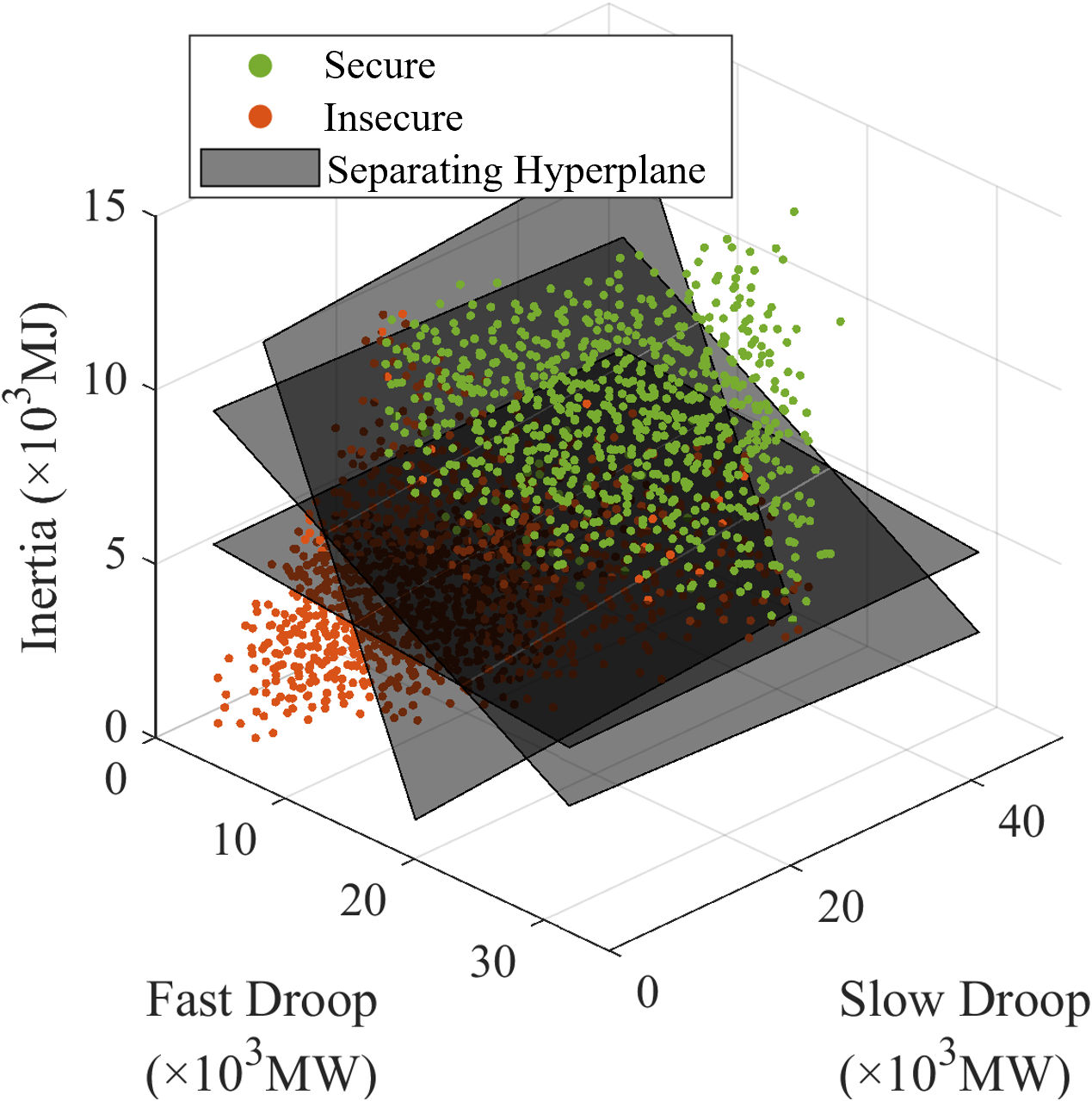}
        \label{fig:dataset3}}
\caption{Results of dataset generation and frequency constraint extraction.}
\label{fig:dataset_generation}
\end{figure}

The simulation-based SFR dataset is generated following the methodology described in Section \ref{section:dataset_generation}. 
Fig. \ref{fig:dataset_generation} illustrates the results of dataset generation and frequency constraint extraction.
First, an original dataset is obtained by running conventional unit commitment model or from the historical system operation records.
Equivalent aggregated parameters are computed from these data, as visualized in Fig. \ref{fig:dataset0}. 
These parameters primarily depend on online generator combinations, resulting in relatively concentrated distributions due to limited variations in unit commitment patterns.
To enhance dataset diversity, random perturbations are introduced to key SFR parameters, followed by selection of representative data points (Fig. \ref{fig:dataset1}).

To rigorously capture the coupling effects of emergency controls and power imbalances, for each operational data point, we conducted simulations considering 10 randomized EPC schemes, 10 randomized DLC schemes, and 40 different power imbalance scenarios.
The final dataset comprises three key feature categories: equivalent aggregated parameters, power imbalance magnitudes, and EFC schemes. 
Fig. \ref{fig:dataset2} specifically illustrates a three-dimensional projection of the labeled dataset for the case with $\Delta P_{D,a}=200MW$, $\Delta P_{EPC,a}=50MW$, and $\Delta P_{DLC,a}=50MW$.

To ensure the weighted oblique decision tree (WODT) focuses on the critical security boundary, we specifically set the frequency security threshold as 0.5 Hz and selected data points with maximum frequency deviations in the range of 0.4 Hz to 0.6 Hz. This selective sampling strategy improves the classifier's precision near the 0.5 Hz security threshold. The final statistics of the generated datasets for the three areas are as follows:
\begin{itemize}[]
	\item Area 1: The dataset contains 1,123,210 samples, comprising 42.2\% insecure samples (violation) and 57.8\% secure samples.
	\item Area 2: The dataset contains 1,055,099 samples, comprising 39.9\% insecure samples and 60.1\% secure samples.
	\item Area 3: The dataset contains 1,062,983 samples, comprising 42.2\% insecure samples and 57.8\% secure samples.
\end{itemize}
This large-scale, balanced dataset ensures the robust training of the WODT for accurate frequency constraint extraction.
A WODT is trained on the dataset to capture the frequency dynamics of each area, with the resulting decision rules forming the frequency security constraints.
Fig. \ref{fig:dataset3} demonstrates the classification boundary of a secure leaf node, where the gray separating planes correspond to the linear rules learned by the WODT.

Table \ref{tab:wodt_coefficients} lists the specific coefficients of the splitting hyperplanes learned by the WODT at different depths (for the Area 1 dataset). Each row represents a linear separating boundary at a certain depth of the form $\sum w_i x_i + \text{Bias} \ge 0$. The magnitude and sign of the coefficients reflect the sensitivity of the frequency nadir to different system parameters. For instance, the coefficients for the initial power imbalance ($\Delta P_{D}$) are consistently large and negative, indicating that a larger disturbance reduces the security margin. Conversely, the coefficients for system inertia ($H$) and emergency control resources ($\Delta P_{EPC}$, $\Delta P_{DLC}$) are positive, reflecting their role in supporting frequency stability.
\begin{table*}[htbp]
	\caption{Learned Coefficients of the WODT Hyperplanes (Depth=3)}
	\label{tab:wodt_coefficients}
	\centering
	\setlength{\tabcolsep}{3pt}
	\begin{tabular}{cccccccc}
		\toprule
		\multirow{2}{*}{Node} & $H$ & $D^{fast}$ & $D^{slow}$ & $\Delta P_{EPC}$ & $\Delta P_{DLC}$ & $\Delta P_{D}$ &  \multirow{2}{*}{Bias} \\
		& ($w_1$) & ($w_2$) & ($w_3$) & ($w_4$) & ($w_5$) & ($w_6$) & \\
		\midrule
		Depth=1 & $8.68 \times 10^2$ & $4.70 \times 10^2$ & $2.53 \times 10^1$ & $9.25 \times 10^2$ & $7.06 \times 10^2$ & $-1.87 \times 10^3$ & $3.08$ \\
		
		Depth=2 & $2.30 \times 10^3$ & $1.77 \times 10^2$ & $7.76 $ & $9.24 \times 10^2$ & $2.96 \times 10^2$ & $-2.21 \times 10^3$ & $2.84 \times 10^1$ \\
		
		Depth=3 & $1.82 \times 10^1$ & $3.83 \times 10^1$ & $3.40$ & $4.27 \times 10^1$ & $3.98 \times 10^1$ & $-8.09 \times 10^1$ & $1.47$ \\
		\bottomrule
		\multicolumn{8}{l}{\footnotesize Note: The hyperplane equation is $\sum_{i=1}^{6} w_i x_i + \text{Bias} \ge 0$.} \\
	\end{tabular}
\end{table*}

To systematically validate the accuracy of the WODT and determine the optimal depth, we iteratively train the WODT with incremental depth and compare the accuracy with alternative classification methods, including the linear support vector machine (Linear SVM) and the neural network (NN). The results are presented in Table \ref{table:accuracy}.
The benchmark NN architecture employs ReLU activation and two hidden layers (128 and 64 neurons respectively).
The accuracy of the WODT improves substantially as the depth increases from 1 to 3, and the accuracy improvement saturates when the depth exceeds 3.
At the depth of 3, WODT achieves comparable accuracy to NN while significantly outperforming Linear SVM.
Crucially, WODT implementation avoids the computational burden associated with NN-based constraint embedding, offering an efficient balance between accuracy and solution tractability.
\begin{table}[htbp]
\centering
\setlength{\tabcolsep}{9pt}
\caption{Classification Accuracy of Different Methods}
\label{table:accuracy}
\begin{tabular}{ccccc}
\toprule
\multicolumn{2}{c}{\multirow{2}{*}{Method}} & \multicolumn{3}{c}{Accuracy} \\ \cmidrule(l){3-5} 
\multicolumn{2}{c}{}                        & Area 1   & Area 2  & Area 3  \\ \midrule
\multirow{6}{*}{WODT}       & Depth=1       & 97.08\%  & 97.12\% & 96.82\% \\
                            & Depth=2       & 98.35\%  & 98.17\% & 98.52\% \\
                            & \textbf{Depth=3}       & \textbf{99.03\%}  & \textbf{98.82\%} & \textbf{99.16\%} \\
                            & Depth=4       & 99.03\%  & 98.83\% & 99.16\% \\
                            & Depth=5       & 99.06\%  & 98.83\% & 99.18\% \\
                            & Depth=6       & 99.06\%  & 98.87\% & 99.19\% \\ \midrule
\multicolumn{2}{c}{Linear SVM}              & 97.00\%  & 96.99\% & 96.94\% \\ \midrule
\multicolumn{2}{c}{NN}                      & 99.36\%  & 99.18\% & 99.35\% \\ \bottomrule
\end{tabular}
\end{table}

\subsection{Planning Results}
The proposed emergency-aware and frequency-constrained HVDC planning method is evaluated through comparison of three different cases, namely:
\begin{itemize}
    \item Non-FC: The conventional planning model without frequency constraints.
    \item FC: The planning model considering frequency constraints, while the coordinated EFC is disabled.
    \item \textbf{FC-EC}: The proposed planning model considering frequency constraints, with the coordinated EFC enabled.
\end{itemize}

Table \ref{install} details the installed HVDC lines and energy storage systems, while Table \ref{optim_result} compares the optimization results. 
In the non-FC case, two HVDC lines are installed to transmit power from Area 3 to Area 1 and Area 2 with the capacity of 800MW and 350 MW respectively, achieving the lowest total cost through economies of scale.
However, this configuration results in frequency deviations exceeding security limits during HVDC fault emergencies due to the large active power imbalance.
In practice, large-scale load shedding will be triggered before the system frequency reaches the value simulated by the SFR model.
Both FC and FC-EC cases maintain frequency deviations within the security bound. 
In the FC case, only local primary frequency regulation resources can be exploited to maintain the frequency security.
Consequently, the capacity of each HVDC line is restricted to 200MW to avoid large power imbalance and the total installed HVDC capacity is small, which hinders the accommodation of renewable energy produced in Area 3 and significantly increases the operational cost.
This conservative approach proves economically inefficient.
In contrast, the FC-EC case achieves near-optimal total HVDC capacity with individual lines limited within 400MW.
Leveraging emergency control resources reduces total costs by 6.77\% compared to the FC case while maintaining frequency security.
More energy storage systems are installed to enhance frequency regulation due to the increased individual HVDC capacity.
Overall, the FC-EC case effectively addresses both frequency security and cost-effectiveness considerations, providing the most balanced and practical planning scheme.

\begin{table}[htbp]
\centering
\setlength{\tabcolsep}{4pt}
\caption{The installed generators and HVDC lines}
\label{install}
\begin{tabular}{ccc}
\toprule
Model & Installed HVDC & Installed ESS\\ 
\midrule
non-FC  & \begin{tabular}[c]{c}HVDC 4 (Area 3 to Area 2): 800.0MW\\      HVDC 6 (Area 3 to Area 1): 350.0MW\end{tabular}                        & ESS 9 (Area 3)                                                                                            \\ \midrule
FC    & \begin{tabular}[c]{c}HVDC 3 (Area 3 to Area 2): 200.0MW\\HVDC 4 (Area 3 to Area 2): 200.0MW\\      HVDC 6 (Area 3 to Area 1): 200.0MW\\      HVDC 7 (Area 3 to Area 1): 200.0MW\end{tabular} & \begin{tabular}[c]{c}ESS 3 (Area 1)\end{tabular} \\ \midrule
\textbf{FC-EC} & \begin{tabular}[c]{c}HVDC 2 (Area 1 to Area 2): 300.0MW\\      HVDC 3 (Area 3 to Area 2): 350.0MW\\HVDC 4 (Area 3 to Area 2): 350.0MW\\      HVDC 6 (Area 3 to Area 1): 350.0MW\end{tabular} & \begin{tabular}[c]{c}ESS 7 (Area 3)\\ ESS 8 (Area 3)\\ ESS 9 (Area 3)\end{tabular}                                                                                            \\ \bottomrule
\end{tabular}
\end{table}

\begin{table}[htbp]
\centering
\caption{The optimization result comparison}
\label{optim_result}
\setlength{\tabcolsep}{2pt}
\begin{tabular}{@{}ccccccc@{}}
\toprule
\multirow{2}{*}{Model} & \multirow{2}{*}{\begin{tabular}[c]{@{}c@{}}Total \\ Cost (M\$)\end{tabular}} & \multirow{2}{*}{\begin{tabular}[c]{@{}c@{}}Investment \\ Cost (M\$)\end{tabular}} & \multirow{2}{*}{\begin{tabular}[c]{@{}c@{}}Operational \\ Cost (M\$)\end{tabular}} & \multicolumn{3}{c}{\begin{tabular}[c]{@{}c@{}}Maximum Frequency \\ Deviation (Hz)\end{tabular}} \\ \cmidrule(l){5-7} & & & & Area 1 & Area 2 & Area 3 \\ \midrule 
non-FC & 715.13 & 37.81 & 677.32 & 2.24 & 4.89 & 1.66 \\ 
\midrule
FC & 779.52 & 48.60 & 730.92 & 0.47 & 0.38 & 0.49 \\ \midrule
\textbf{FC-EC} & 726.71 & 60.21 & 666.50 & 0.49 & 0.50 & 0.41 \\ \bottomrule
\end{tabular}
\end{table}

To further validate the necessity of the proposed data-driven approach, we compared the planning results with a traditional model-based method, specifically, the low-order system frequency response (LOSFR) model with piecewise linearization for constraint extraction \cite{ahmadi2014security}. While the LOSFR model is widely used due to its computational simplicity, it ignores small time constants and complex control effects (e.g., the specific dynamics of water hammer effects).

Table \ref{tab:lsfr_comparison} presents the comparison results. It is observed that the LOSFR-based method yields a planning scheme with slightly lower total costs compared to the proposed method. However, this cost reduction comes at the expense of system security. As shown in the frequency deviation columns, the plans generated by the LOSFR method result in frequency deviations significantly exceeding the 0.5 Hz security threshold, especially in the FC-EC case.
This indicates that the simplified LOSFR model underestimates the severity of frequency nadirs. In contrast, the proposed WODT approach accurately captures the complex non-linear frequency dynamics embedded in the high-fidelity simulation data.

\begin{table}[htbp]
	\centering
	\caption{Comparison with LOSFR-Based Method}
	\label{tab:lsfr_comparison}
	\setlength{\tabcolsep}{4pt}
	\begin{tabular}{cccccc}
		\toprule
		\multirow{2}{*}{Model} &
		\multirow{2}{*}{\begin{tabular}[c]{@{}c@{}}Constraint \\ Method \end{tabular}} & \multirow{2}{*}{\begin{tabular}[c]{@{}c@{}}Total \\ Cost (M\$)\end{tabular}} & \multicolumn{3}{c}{\begin{tabular}[c]{@{}c@{}}Maximum Frequency \\ Deviation (Hz)\end{tabular}} \\ \cmidrule(l){4-6} & & & Area 1 & Area 2 & Area 3 \\
		\midrule
		\multirow{2}{*}{FC} & LOSFR &772.50 & 0.63 & 0.59 & 0.52 \\ \cmidrule{2-6}
		 & Proposed &779.52 & 0.47 & 0.38 & 0.49 \\ \midrule
		\multirow{2}{*}{FC-EC} &LOSFR & 725.65 & 0.66 & 0.72 & 0.51 \\ \cmidrule{2-6}
		 &Proposed & 726.71 & 0.49 & 0.50 & 0.41 \\ \bottomrule
	\end{tabular}
\end{table}

\begin{figure}[htbp]
    \centering
    \includegraphics[width=0.48\textwidth]{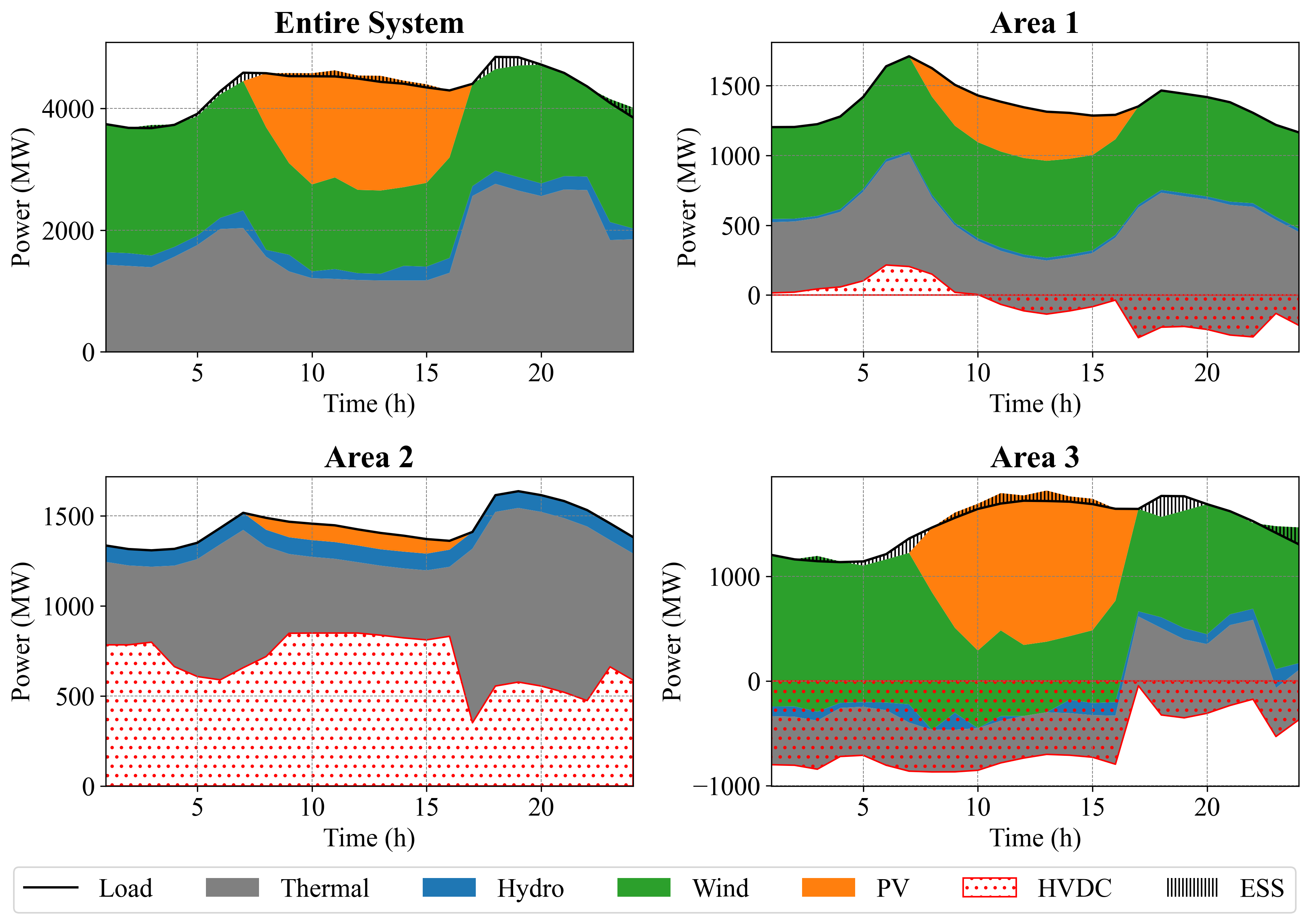}
    \caption{Power generation portfolio.}
    \label{fig:stack}
\end{figure}

\begin{figure}[htbp]
    \centering
    \includegraphics[width=0.38\textwidth]{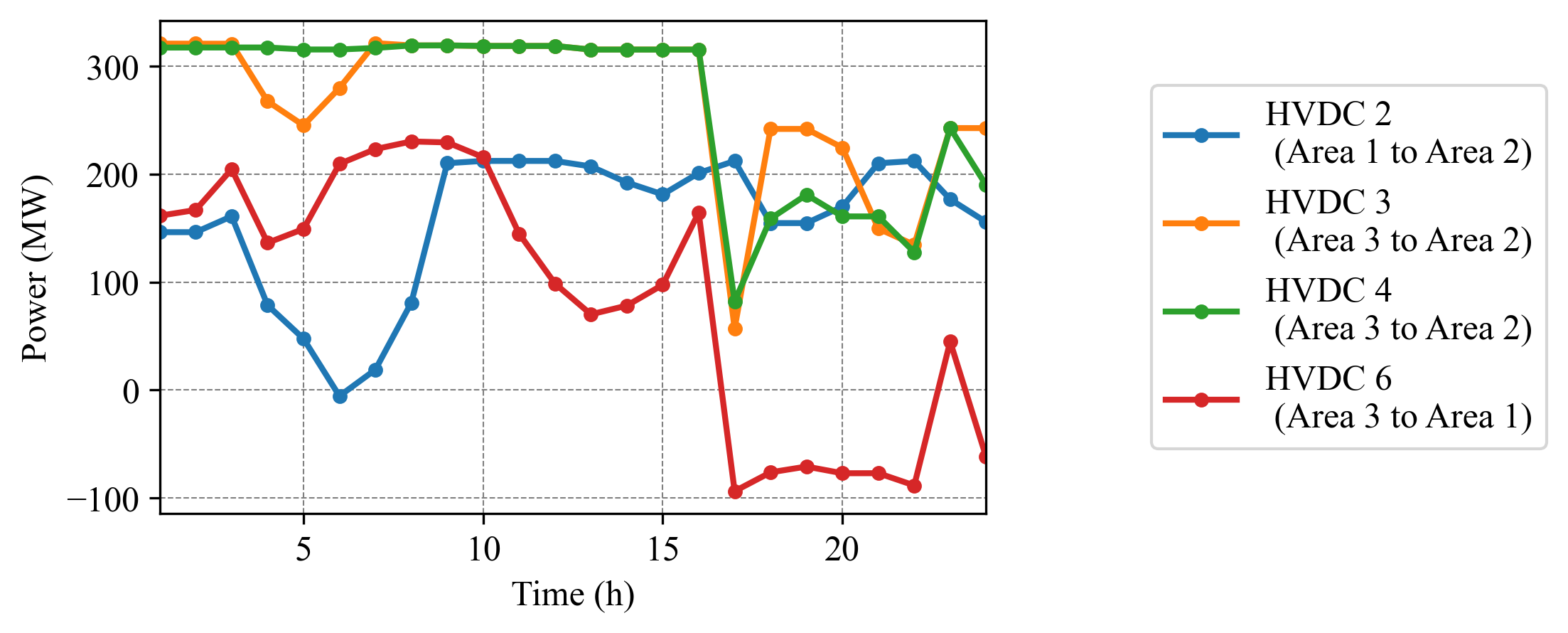}
    \caption{Active power transferred by each HVDC lines.}
    \label{fig:hvdc}
\end{figure}

We choose the representative scenario with the highest probability as an example to demonstrate the system operational characteristics with the planning scheme of the FC-EC case.
Fig. \ref{fig:stack} presents the power generation portfolio of the whole system and all the areas, while Fig. \ref{fig:hvdc} presents the active power transferred by each HVDC lines.
Area 3 is a sending-end grid with substantial PV and wind capacity, typically exporting power to Areas 1 and 2 via HVDC 3, 4, and 6. 
During nighttime periods without PV output, the net power export of Area 3 is low, and the power flow between Area 3 and Area 1 (HVDC 6) may reverse direction.
Area 2 is a typical receiving-end grid, with a large proportion of load served by the power imported from Area 1 and Area 3.

\begin{figure}[htbp]
    \centering
    \includegraphics[width=0.48\textwidth]{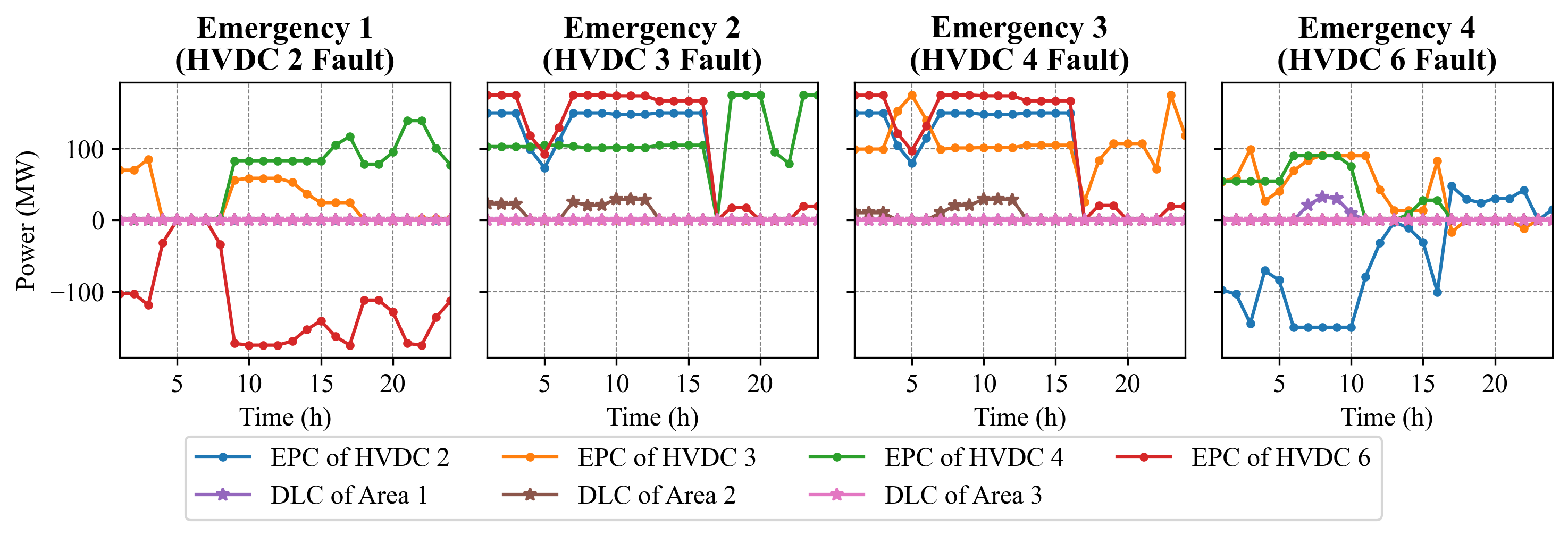}
    \caption{The response power of HVDC EPC and DLC under different emergencies.}
    \label{fig:ec}
\end{figure}

We further investigate into the mechanism of the coordinated emergency frequency control scheme.
Fig. \ref{fig:ec} presents the response power of HVDC EPC and DLC under different emergencies, with each emergency representing an HVDC fault.
During HVDC 2 (Area 1 to Area 2) fault, HVDC 3 and HVDC 4 increase the power transmission from Area 3 to Area 2 to mitigate the power shortage of Area 2, while HVDC 6 decreases the power transmission from Area 3 to Area 1 to mitigate the power surplus of Area 1.
During HVDC 3 (Area 3 to Area 2) fault, HVDC 4 increases its power to directly undertake part of the pre-fault power of HVDC 3.
Meanwhile, HVDC 6 increases the power transmission from Area 3 to Area 1 and HVDC 2 increases the power transmission from Area 1 to Area 2, together forming an alternative power flow path from Area 3 to Area 2.
Concurrently, the direct load control of Area 2 responds to further mitigate the power shortage.
The situation during HVDC 4 fault is very similar, since HVDC 4 is parallel with HVDC 3.
In the case of HVDC 6 (Area 3 to Area 1) fault, HVDC 3 and HVDC 4 increase the power exported from Area 3 and HVDC 2 decreases the power exported from Area 1.
For certain time periods, the direct load control of Area 1 is utilized to restrain the frequency drop.
The circular HVDC network configuration among asynchronously interconnected areas takes advantage of HVDC emergency control and proves effective to provide mutual frequency support during HVDC faults.

\begin{figure}[htbp]
    \centering
    \includegraphics[width=0.48\textwidth]{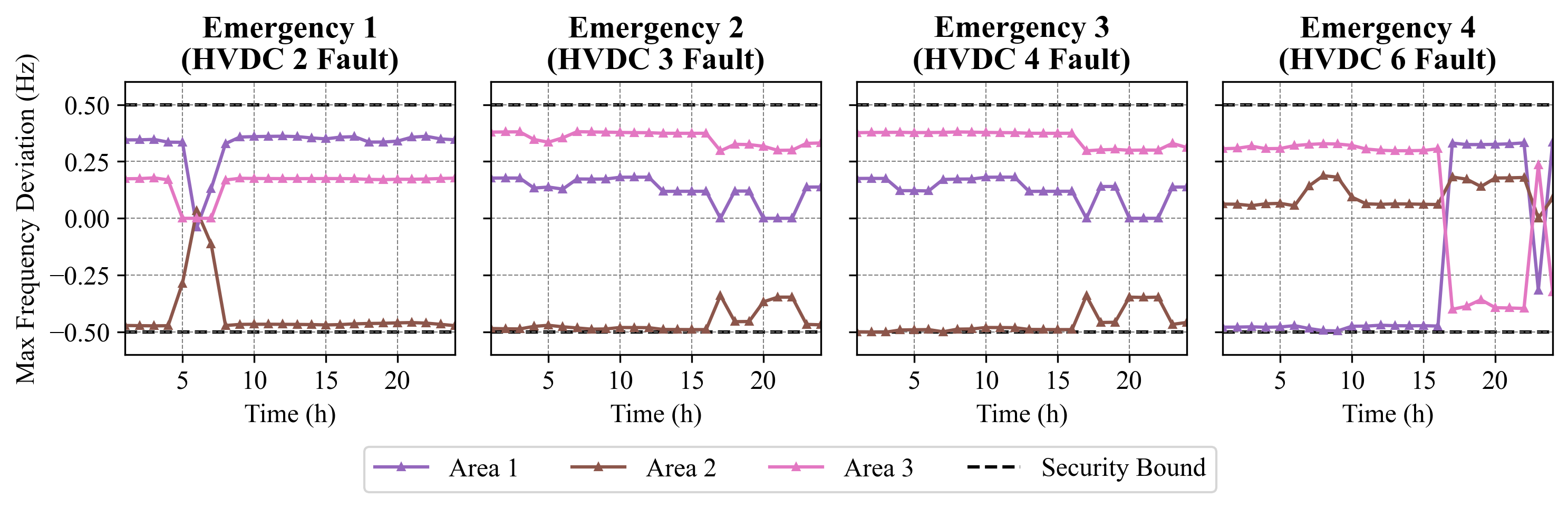}
    \caption{The maximum frequency deviations of each area.}
    \label{fig:ec_deltaf}
\end{figure}

Fig. \ref{fig:ec_deltaf} presents the maximum frequency deviations of each area.
During HVDC fault emergencies, the sending-end grid experiences a frequency rise and the receiving-end grid experiences a frequency drop, but both can be controlled effectively within the ±0.5 Hz security bound.
The area not directly connected to the fault HVDC can also be influenced by HVDC EPC, but the frequency fluctuation is typically below 0.2 Hz, demonstrating the system's ability to localize disturbance impacts while preserving overall network security.

\subsection{Sensitivity Analysis on Emergency Control Parameters}

To assess the robustness of the proposed planning method, we conducted a sensitivity analysis on the economic and technical parameters of the emergency control layer. Specifically, we examined the impact of variations in the unit control costs ($C_{DLC}$ and $C_{EPC}$) and the maximum available Direct Load Control (DLC) resources ($\Delta P_{DLC}^{max}$). It should be noted that the maximum available resource for HVDC EPC was not treated as an independent variable in this analysis. Unlike DLC, which is constrained by external load characteristics, the available EPC capacity is intrinsically determined by the installed capacity of the HVDC tie-lines. Since the HVDC installed capacity is a decision variable optimized within the planning model, the EPC limit is an endogenous variable rather than an exogenous parameter suitable for independent sensitivity testing.

Table \ref{tab:sensitivity} summarizes the planning results under six sensitivity cases compared to the baseline (FC-EC case). The baseline parameters are defined as $C_{DLC}=\$1000/MW$, $C_{EPC}=\$100/MW$, and the DLC resource limit is set to 2\% of the total load. In Cases 1 through 4, the control costs are varied by $\pm50\%$ relative to the baseline values. In Cases 5 and 6, the available DLC resource capacity is varied by $\pm50\%$. 

The results indicate that the proposed planning framework adapts to these variations while maintaining frequency security. 
The planning results (HVDC and ESS capacities) remain unchanged in Cases 1 through 4. This implies that the installation of flexibility resources is driven by technical frequency security constraints rather than economic parameters between investment and emergency operation costs. Since HVDC faults are low-probability events, the expected value of the operational penalty is not sufficient to influence capital-intensive infrastructure decisions.

The model is sensitive to the physical availability of emergency resources. In Case 5, where the available DLC resource is reduced by 50\%, the system faces a flexibility deficit. Consequently, the model increases the ESS capacity to 200 MW (an additional 50 MW unit) to provide the necessary fast frequency response. Conversely, increasing the DLC limit (Case 6) does not reduce the ESS capacity, although the total cost decreases slightly. This implies that the additional DLC capacity is not fully utilized during emergencies, primarily due to its higher operational cost compared to other resources.

\begin{table}[htbp]
	\caption{Sensitivity Analysis of Control Costs and Resource Availability}
	\label{tab:sensitivity}
	\centering
	\setlength{\tabcolsep}{2pt}
	\begin{tabular}{ccccccc}
		\toprule
		\multirow{2}{*}{Case} & \multirow{2}{*}{$C_{DLC}$} & \multirow{2}{*}{$C_{EPC}$} & \multirow{2}{*}{$\Delta P_{DLC}^{max}$} & Total HVDC & Total ESS & Total Cost \\
		&&&& Cap. (MW) & Cap. (MW) & (M\$) \\
		\midrule
		Baseline & 100\% & 100\% & 100\% & 1350 & 150 & 726.71 \\
		Case 1 & 150\% & 100\% & 100\% & 1350 & 150 & 726.92 \\
		Case 2 & 50\% & 100\% & 100\% & 1350 & 150 & 726.34 \\
		Case 3 & 100\% & 150\% & 100\% & 1350 & 150 & 726.85 \\
		Case 4 & 100\% & 50\% & 100\% & 1350 & 150 & 726.47 \\
		Case 5 & 100\% & 100\% & 50\% & 1350 & 200 & 727.12 \\
		Case 6 & 100\% & 100\% & 150\% & 1350 & 150 & 726.61 \\
		\bottomrule
	\end{tabular}
\end{table}

\section{Conclusion}
\label{conclusion}

This paper presents an emergency-aware and frequency-constrained HVDC planning method for optimizing inter-area HVDC capacity in a multi-area asynchronously interconnected grid. 
A coordinated emergency frequency control scheme is developed for HVDC faults to optimally allocate frequency regulation resources, including direct load control and HVDC emergency power control.
An enhanced system frequency response model incorporating event-driven emergency frequency control is established, with equivalent parameter aggregation employed to reduce model dimensionality. 
Frequency nadir constraints are extracted from a simulation-generated dataset using a weighted oblique decision trees approach.
Formulated as a stochastic program, the proposed planning model accounts for diverse scenarios and potential emergencies.
Case study validates the effectiveness of the emergency frequency control and the accuracy of the frequency constraints. 
The planning results show significant economic advantages while maintaining system frequency security under all possible HVDC fault emergencies. 
The proposed planning model is useful to balance cost efficiency with frequency security requirements, providing a practical solution for inter-area HVDC planning.





\ifCLASSOPTIONcaptionsoff
  \newpage
\fi



\bibliographystyle{IEEEtran}
\bibliography{refs.bib}
%

%

\begin{IEEEbiography}[{\includegraphics[width=1in,height=1.25in,clip,keepaspectratio]{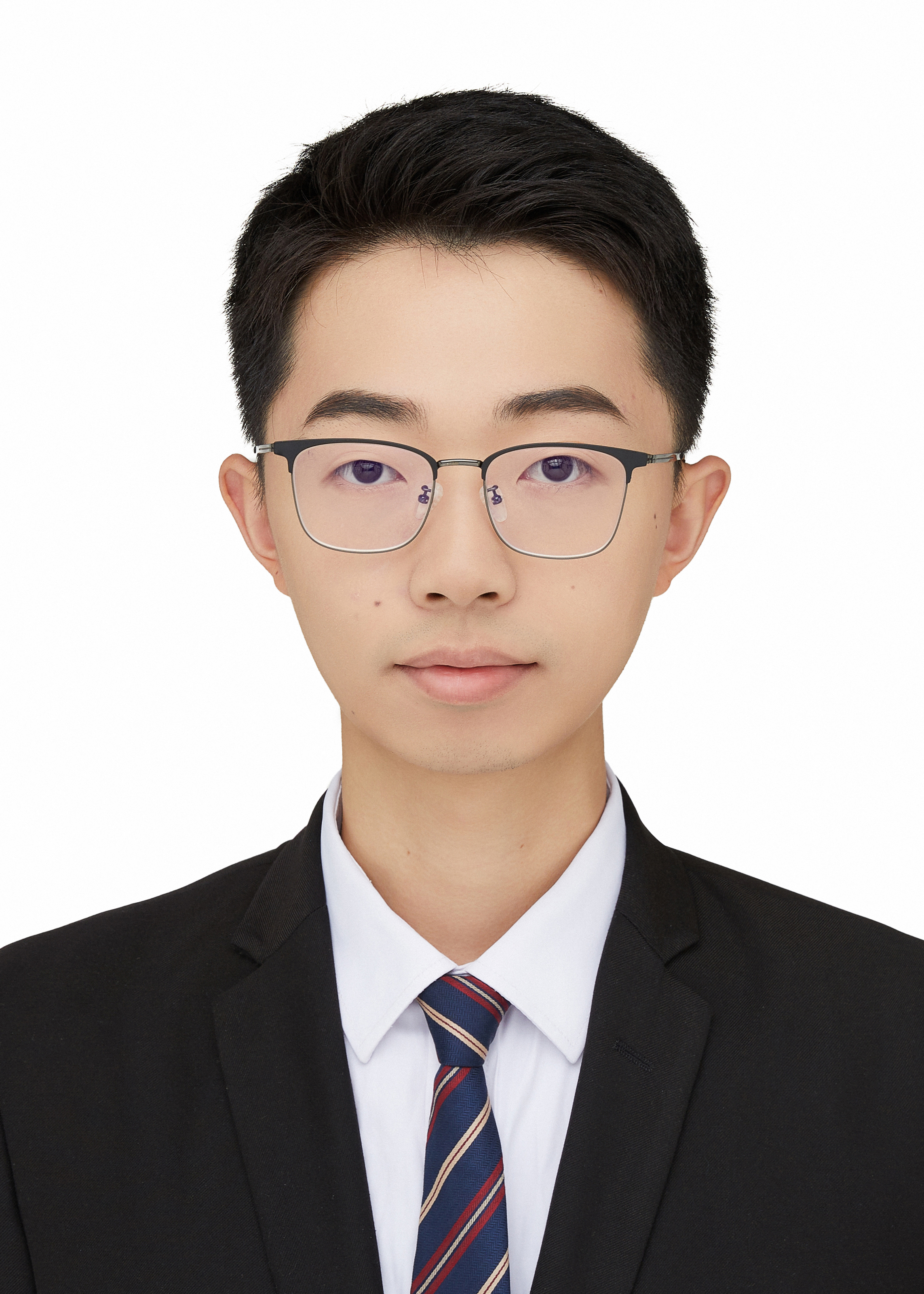}}]{Yiliu He}
	(Student Member, IEEE) received the B.S. degree in electrical engineering from Tsinghua University, Beijing, China, in 2021, where he is currently pursuing the Ph.D. degree. His research interests include stability-constrained power system optimization, demand response, and application of data-driven methods in power systems.
\end{IEEEbiography}
\begin{IEEEbiography}[{\includegraphics[width=1in,height=1.25in,clip,keepaspectratio]{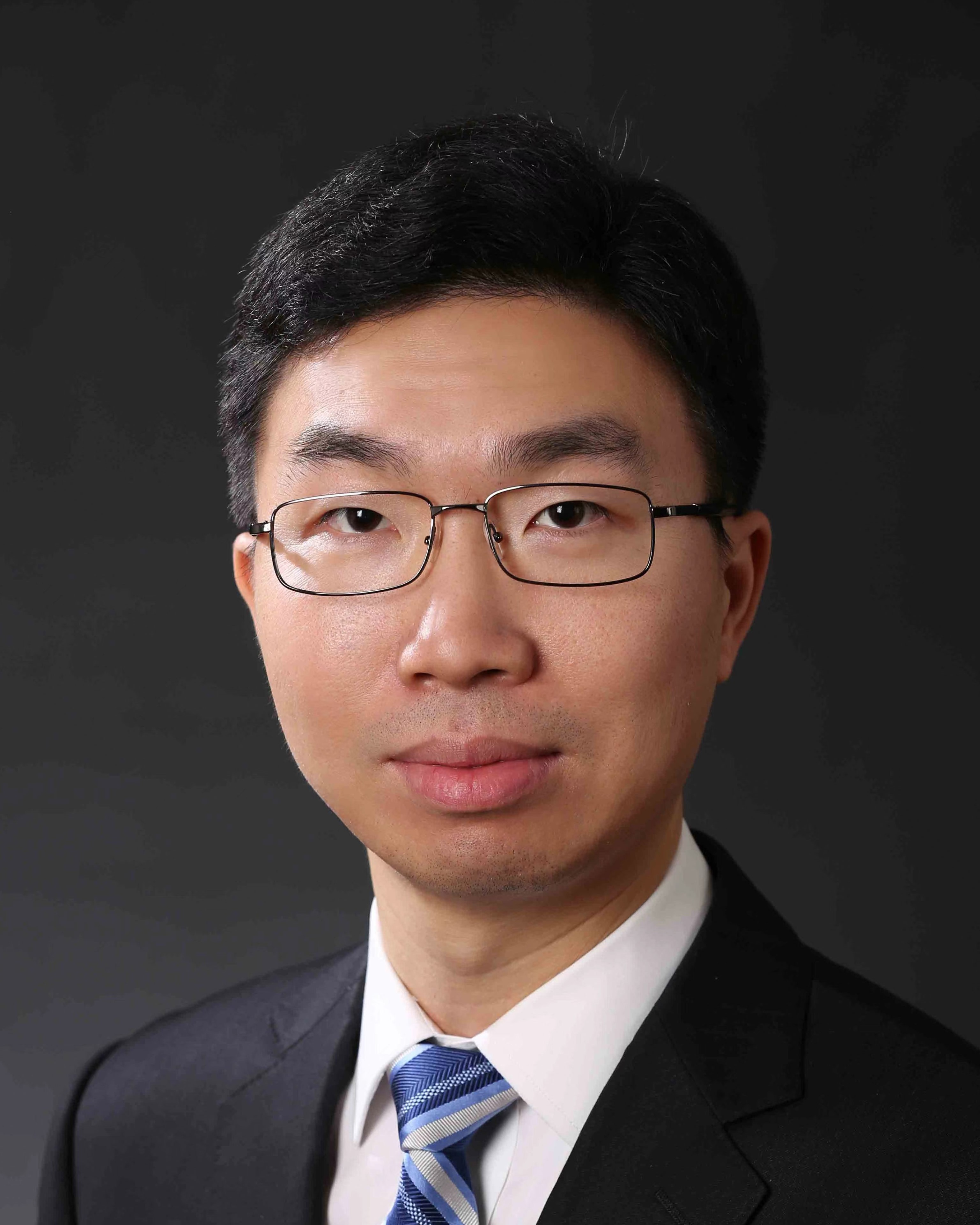}}]{Haiwang Zhong}
	(IEEE Senior Member) received the B.S. and Ph.D. degrees in electrical engineering from Tsinghua University.
	He is currently an Associate Professor with the Department of Electrical Engineering, Tsinghua University. He is also the Director of Energy Internet Trading and Operation Research Department of Sichuan Energy Internet Research Institute, Tsinghua University. His research interests include power system operations and optimization, electricity markets. He was a recipient of the ProSPER.Net Young Scientist Award and the Outstanding Postdoctoral Fellow of Tsinghua University. He serves as an Associate Editor for the CSEE Journal of Power and Energy Systems. 
	He currently serves as the Convenor of CIGRE C1.54 Working Group on Assessment of system reserves and flexibility needs in the power systems of the future.
\end{IEEEbiography}
\begin{IEEEbiography}[{\includegraphics[width=1in,height=1.25in,clip,keepaspectratio]{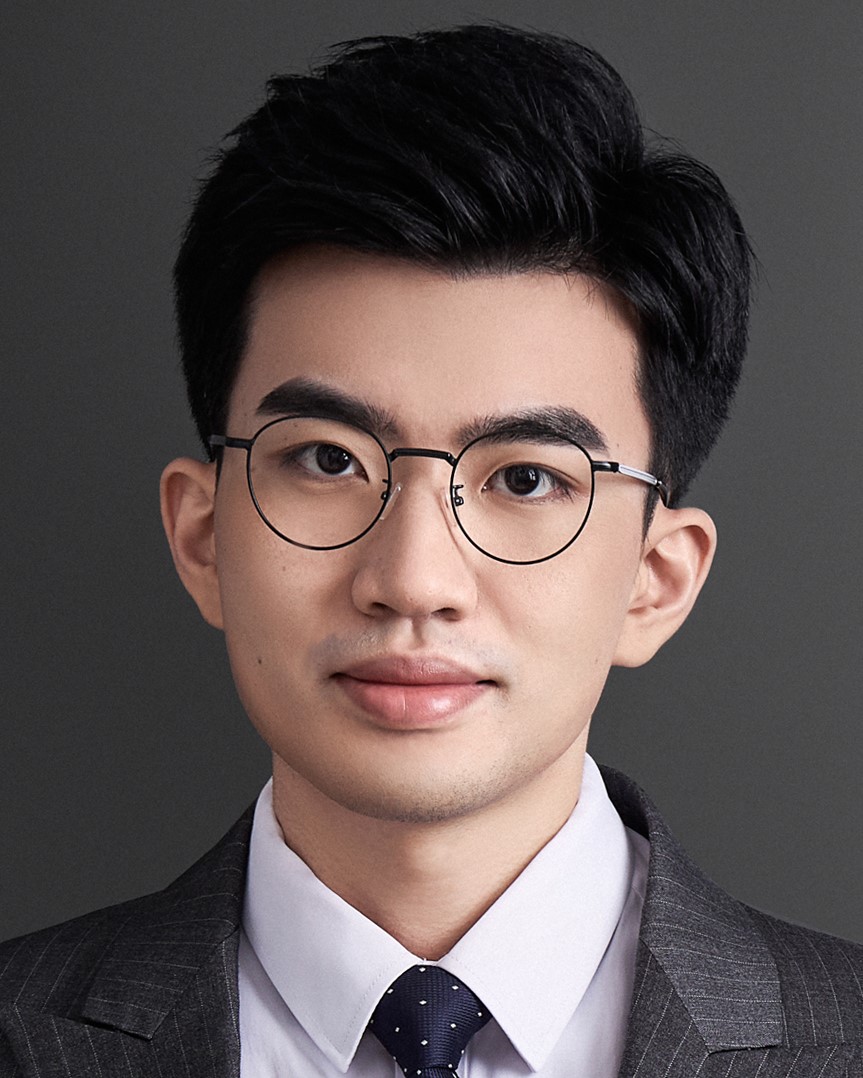}}]{Grant Ruan}
	(Member, IEEE) is currently a postdoc with the Laboratory for Information \& Decision Systems (LIDS) at MIT. Before joining MIT, he received the Ph.D. degree in electrical engineering from Tsinghua University in 2021, and worked as a postdoc with the University of Hong Kong in 2022. He visited Texas A\&M University in 2020 and the University of Washington in 2019. His research interests include electricity market, energy resilience, demand response, data science and machine learning applications.
\end{IEEEbiography}

\begin{IEEEbiography}[{\includegraphics[width=1in,height=1.25in,clip,keepaspectratio]{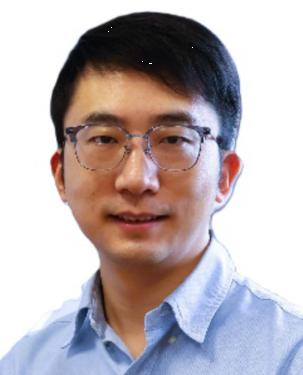}}]{Yan Xu}
	(Senior Member, IEEE) received the B.E. and M.E. degrees in electrical engineering from the South China University of Technology, Guangzhou, China, in 2008 and 2011, respectively, and the Ph.D. degree in electrical engineering from the University of Newcastle, Callaghan NSW, Australia, in 2013.
	
	After the postdoctoral training with the University of Sydney Postdoctoral Fellowship, he joined Nanyang Technological University (NTU) with the Nanyang Assistant Professorship. He was an Associate Professor in 2021, and the Cham Tao Soon Professor in Engineering (an endowed professorship named after the founding president of NTU) in 2024. He is currently the Director with Center for Power Engineering (CPE), and Co-Director with SingaporePower Group-NTU Joint Lab, with NTU. His research interests include power system stability and control, microgrids, and data-analytics for smart grid applications.
	Dr Xu was the recipient of the Nanyang Research Award (Young Investigator) by NTU, and the Outstanding Engineer Award by IEEE Power \& Energy Society (PES) Singapore Chapter. He is an Associate Editor for IEEE TRANSACTIONS ON SMART GRID and IEEE TRANSACTIONS ON POWER SYSTEMS, Chair of IEEE PES Singapore Chapter (2021 and 2022), and the General Co-Chair of the 11th IEEE ISGT-Asia Conference in 2022.
\end{IEEEbiography}

\begin{IEEEbiography}[{\includegraphics[width=0.95in,height=1.25in,clip]{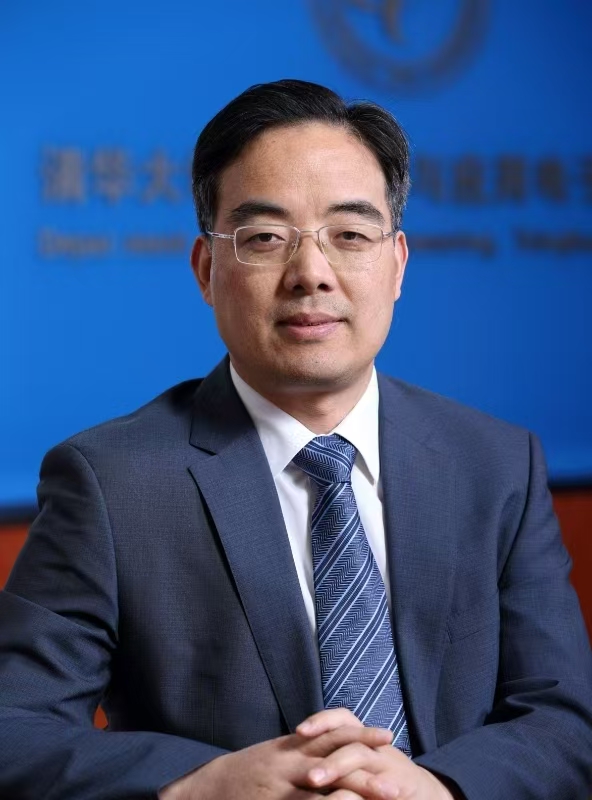}}]{Chongqing Kang}
(M'01-SM'07-F'17) received his Ph.D. degree from the Department of Electrical Engineering, Tsinghua University, in 1997.  

He is now a Professor with the Department of Electrical Engineering, Tsinghua University. His research interests include power system planning, power system operation, renewable energy, low carbon electricity technology and load forecasting. 
\end{IEEEbiography}








\end{document}